\documentclass{article}
\usepackage{graphicx}

\usepackage{jheppub}
\usepackage{amsmath}
\usepackage{epsfig,multicol,bbm}
\usepackage{url}

\usepackage{graphicx}
\usepackage{setspace}
\usepackage{amssymb}
\usepackage{listings}
\usepackage{epstopdf}
\usepackage{float}
\usepackage{fancyheadings}
\usepackage{color}
\usepackage{amsmath}
\usepackage{ dsfont }
\usepackage{slashed}
\usepackage{subfigure}

\newcommand{\newc}{\newcommand}
\newc{\gsim}{\lower.7ex\hbox{$\;\stackrel{\textstyle>}{\sim}\;$}}
\newc{\lsim}{\lower.7ex\hbox{$\;\stackrel{\textstyle<}{\sim}\;$}}
\newc{\gev}{\,{\rm GeV}}
\newc{\mev}{\,{\rm MeV}}
\newc{\ev}{\,{\rm eV}}
\newc{\kev}{\,{\rm keV}}
\newc{\tev}{\,{\rm TeV}}

\newcommand{\ifb}{\,\mathrm{fb}^{-1}}

\newc{\mz}{M_Z}
\newc{\mpl}{M_*}
\newc{\mw}{m_{\rm weak}}
\newc{\nr}[1]{N^c_R{}_{#1}}
\usepackage{amsmath}

\def\slash#1{\setbox0=\hbox{$#1$}#1\hskip-\wd0\hbox to\wd0{\hss\sl/\/\hss}}

\def\beq{\begin{equation}}
\def\eeq{\end{equation}}
\def\bea{\begin{eqnarray}}
\def\eea{\end{eqnarray}}
\def\bitem{\begin{itemize}}
\def\eitem{\end{itemize}}
\newc{\ie}{{\it i.e.}}          \newc{\etal}{{\it et al.}}
\newc{\eg}{{\it e.g.}}          \newc{\etc}{{\it etc.}}
\newc{\cf}{{\it c.f.}}

\def\bar#1{\overline{#1}}

\def\vector#1{{\vec{#1}}}
\def\inv{^{\raise.15ex\hbox{${\scriptscriptstyle -}$}\kern-.05em 1}}
\def\lbar{{\lower.35ex\hbox{$\mathchar'26$}\mkern-10mu\lambda}} %lambda bar

\def\to{\rightarrow}

\let\<=\langle
\let\>=\rangle

\let\+=\uparrow
\newcommand\fverb{\setbox\fverbbox=\hbox\bgroup\verb}
\newcommand\fverbdo{\egroup\medskip\noindent%
			\fbox{\unhbox\fverbbox}\ }
\newcommand\fverbit{\egroup\item[\fbox{\unhbox\fverbbox}]}
\newbox\fverbbox

\title{Supersymmetry in the shadow of photini}

\date{\today}

\author[a]{Masha Baryakhtar,}

\author[b,c]{Nathaniel Craig,}

\author[a]{Ken Van Tilburg}

\affiliation[a]{Department of Physics, Stanford University \\
Stanford, CA 94306}
\affiliation[b]{Department of Physics, Rutgers University \\
Piscataway, NJ 08854 }
\affiliation[c]{ School of Natural Sciences, Institute for Advanced Study \\
Princeton, NJ 08540}

\preprint{RUNHETC-2012-11, SU-ITP-12/16}	% OR:
\abstract{Additional neutral gauge fermions -- ``photini'' --  arise in string compactifications as superpartners of $U(1)$ gauge fields.  Unlike their vector counterparts, the photini can acquire
  weak-scale masses from soft SUSY breaking and lead to observable
  signatures at the LHC through mass mixing with the bino.  In this work we
  investigate the collider consequences of adding photini to the neutralino
  sector of the MSSM. Relatively large mixing of one or
  more photini with the bino can lead to prompt decays of the lightest
  ordinary supersymmetric particle; these extra cascades transfer most
  of the energy of SUSY decay chains into Standard Model particles,
  diminishing the power of missing energy as an experimental handle
  for signal discrimination.  We demonstrate that the missing energy in SUSY
  events with photini is reduced dramatically for 
  supersymmetric spectra with MSSM neutralinos near the weak scale, and
  study the effects on limits set by the leading
  hadronic SUSY searches at ATLAS and CMS.  We
  find that in the presence of even one light photino the limits on
  squark masses from hadronic searches can be reduced by $400 \gev$, with
  comparable (though more modest) reduction of gluino mass limits.  We
  also consider potential discovery channels such as dilepton and multilepton searches, which remain sensitive to SUSY
  spectra with photini and can provide an unexpected route to the
  discovery of supersymmetry. Although presented in the context of photini, our results apply in general to theories in which additional light neutral fermions mix with MSSM gauginos.
  }

\begin{document} 

\maketitle

\section{Introduction}

Thus far no convincing evidence for supersymmetry (SUSY) has been discovered at the LHC. While this by no means excludes supersymmetric extensions of the Standard Model, it begins to challenge the most natural and minimal realizations of supersymmetry as an explanation for the hierarchy between the weak scale and Planck scale. On the other hand, there are compelling hints at both ATLAS and CMS \cite{ATLAS:2012ae, Chatrchyan:2012tx} of a new higgs-like state near 125 GeV with approximately Standard Model-like couplings. Should these hints persist and this state prove to be an elementary scalar, supersymmetry remains the most natural explanation for the hierarchy problem consistent with direct and indirect data. 

There are a variety of ways in which a supersymmetric explanation of the weak scale may be reconciled with current LHC limits. One possibility is that the entire spectrum of superpartners is moderately tuned, such that sparticles are not yet kinematically accessible at the LHC; perhaps a percent-level (or worse \cite{ArkaniHamed:2004fb, Giudice:2004tc}) tuning is palatable to Nature. Another possibility is that the mass hierarchy of scalars is inverted relative to the mass hierarchy of fermions, such that only third-generation scalars are light \cite{Dimopoulos:1995mi}; in this case direct production limits remain weak but supersymmetric naturalness is robust \cite{Essig:2011qg, Kats:2011qh, Papucci:2011wy, Brust:2011tb}.\footnote{This scenario is still challenged by limits on new contributions to flavor-changing neutral currents (FCNCs), which require a symmetry relating the soft masses of the first two generations; see e.g. \cite{Giudice:2008uk, Craig:2011yk, Barbieri:2011ci, Craig:2012di}. A related 
possibility is for 
color octet fermions to be relatively heavy, while scalar partners of all three generations are light; this removes one of the primary SUSY production channels and is less constrained by FCNCs, but requires a non-minimal gaugino sector to minimize radiative corrections \cite{Kribs:2012gx}.} 

A third possibility -- and the focus of this work -- is that all
superpartners are light but have thus far evaded detection. The most
important experimental handle in current supersymmetry searches at the
LHC is missing transverse energy ($\slashed{E}_T$) in the final state;
the searches which set the strongest bounds on SUSY parameter space
use $\slashed{E}_T$ or related variables to distinguish signal from
background events
\cite{ref:alphaT1fb,ref:razor5fb,ref:cmsjetsmet1fb,ref:atlas2to6jets5fb}. Missing
energy is a signal of sparticle production if the theory preserves
$R$-parity, in which case the Lightest Supersymmetric Particle (LSP) is
absolutely stable and escapes the detector. While in Standard Model
processes missing energy arises only from neutrinos and energy
mis-measurement, in SUSY processes the lightest stable particle can
carry away a large fraction of the energy in the event as
$\slashed{E}_T$, providing powerful discrimination from Standard Model
backgrounds. Consequently, the strongest limits on SUSY spectra may be
mitigated if the $\slashed{E}_T$ signal is degraded.\footnote{A
  complementary idea is to decrease the amount of {\it visible} energy
  in an event, which exploits the significant visible energy
  requirements in frontier SUSY searches; this may occur when, e.g.,
  the spectrum of states is compressed \cite{LeCompte:2011cn,
    LeCompte:2011fh}.}

One possibility is to relax the assumption of $R$-parity. $R$-parity
is motivated by weak-scale dark matter and baryon number conservation,
but it may be violated in a manner consistent with proton stability;
in this case $\slashed{E}_T$ signatures may be substantially
eroded.\footnote{For recent discussions germane to the LHC, see
  \cite{Carpenter:2007zz,Csaki:2011ge,Dreiner:2012mn,Allanach:2012vj,
    Graham:2012th}.} The disadvantage of $R$-parity violation is that
there is typically no convenient candidate for weak-scale dark matter,
and some care must be taken to ensure proton stability and minimize
flavor violation.

Alternately, $R$-parity may be a good symmetry of the universe, but
sparticle decays still fail to yield substantial $\slashed{E}_T$ due
to additional states in the spectrum beyond the fields of the MSSM. If
one or more of these new states is lighter than the Lightest Ordinary
Supersymmetric Particle (LOSP) of the MSSM, the LOSP is not stable and
subsequently decays to the true LSP. In this case there are various avenues for diminishing the missing energy signals. If the mass splittings in
these new supermultiplets are small, the phase space for decay
products carrying $\slashed{E}_T$ is reduced; this is the mechanism of
``stealth supersymmetry'' \cite{Fan2011, Fan2012}. Stealth SUSY works
particularly well for theories with a low scale of supersymmetry
breaking, where small splittings are natural and additional multiplets
may be motivated by model-building challenges. But another sensible
possibility is simply that the fermionic components of the new degrees
of freedom are relatively light and mix weakly with MSSM states. In
this case, SUSY production at the LHC results in rapid cascade decays
to the LOSP, followed by a prompt or displaced decay into the LSP via
SM states that carry away a significant fraction of the transverse
energy. The diminution of $\slashed{E}_T$ signals in this case is
simply a function of supersymmetric naturalness: the LOSP is close in
mass to Standard Model states such as $W$ and $Z$ bosons or the higgs.

A cynic might worry that such a scenario would merely be proof that Nature is needlessly obfuscatory, and the existence of new light fermions a crueler version of ``who ordered that?'' However, we will argue that there is a highly plausible ultraviolet motivation for precisely this scenario in the form of photini, the gauge fermions of Ramond-Ramond $U(1)$'s that arise from the dimensional reduction of string theory on topologically complex compactification manifolds \cite{Arvanitaki2009}. These new states typically have masses around the mass of the bino and interact with MSSM neutralinos through both kinetic and mass matrix mixing. When this mixing is moderate or large the LOSP can decay promptly to lighter photini, with substantial transverse energy carried off by $Z$ bosons or higgses. The potentially high multiplicity of photini makes this mechanism for reducing $\slashed{E}_T$ well-motivated and reasonably effective at weakening current LHC limits. Although our UV motivation comes from photini, the 
collider phenomenology of additional gauginos mixing with the bino -- and their effects on SUSY searches at the LHC -- are entirely general. In particular, similar effects may arise due to, e.g., goldstinos  \cite{Cheung:2010mc,Cheung:2010qf, Craig:2010yf}, singlinos \cite{Das:2012rr}, or other light $U(1)$ gauge fermions \cite{Abel:2008ai, Ibarra:2008kn, Goodsell:2009xc, Cicoli:2011yh, Goodsell:2011wn}.  

In this paper we illustrate the consequences of light and moderately mixed photini for collider limits on supersymmetry. We begin in Section 2 by reviewing the ultraviolet motivation and infrared phenomenology of photini, as well as their impact in supersymmetric cascade decays. In Section 3 we consider the effects of photini on reducing limits from SUSY searches with the greatest reach, particularly those involving jets and $\slashed{E}_T$. Of course, photini do not entirely extinguish SUSY signatures, and their presence in decay chains may introduce additional hadronic or leptonic activity. In Section 4 we consider the most effective searches for the discovery or exclusion of SUSY with photini, of which the same-sign dilepton and $Z$+jets+$\slashed{E}_T$ searches are particularly promising.

\section{Photini}

We begin by reviewing the ultraviolet motivation for photini from string theory constructions, before turning to the infrared effective theory and the effects on conventional SUSY signals at the LHC. Although our study of LHC phenomenology is motivated by the potential multiplicity of photini arising from string compactifications, it bears emphasizing that the consequences for SUSY searches are generic to any scenario in which additional fermions partially mix with the neutral gauginos of the MSSM, such as singlinos, goldstinos, or additional hidden $U(1)$ gauginos. 

\subsection{Photini in the UV}

Realistic string theory constructions typically result in
extra-dimensional manifolds with rich and non-trivial topology,
including a large number of closed cycles of varying
dimensionality. The topological complexity of the compactification
manifold in turn leaves its imprint on the spectrum of Kaluza-Klein
(KK) zero modes, potentially giving rise to additional light states
beyond those of the MSSM. In particular, the higher-rank tensor fields
intrinsic to string theory give rise to many zero modes upon
compactification, with each zero mode arising from the independent
cycles of the internal manifold. While this is an
entirely generic phenomenon, in our case the most interesting such
tensor fields are the Ramond--Ramond (RR) forms $C_{2,4}$ of type IIB
theory of rank 2 and 4, or the RR forms $C_{1,3}$ of type IIA theory of rank 1 and
3, under which only D-branes are charged.

The statistics of the zero modes resulting from compactification
depend on the class of the cycle and the rank of the tensor field. For
instance, every independent $n$-cycle gives rise to a scalar KK zero
mode in the presence of a rank $n$ form; in certain compactifications
this leads to a plethora of light pseudo-scalar fields with axion-like
couplings \cite{Svrcek:2006yi, Arvanitaki:2009fg}. Similarly, an antisymmetric form of rank $n$ gives
rise also to massless vector fields labeled by the independent cycles
of dimension $(n-1)$. As in the scalar case, these vectors are given
by integrals of the form over the corresponding cycle. For instance,
in type IIB theory each of the 3-cycles $\Sigma^3_i$ allow us to
define a 4d vector field \beq A_\mu^i = \int_{\Sigma^3_i} C_4
\label{U1def}
\eeq by taking three of the four-form indices along the directions of
the cycle. These are truly gauge fields, since they inherit a gauge
symmetry from the underlying Abelian gauge symmetry of the RR field
$C_4\to C_4 +d \Lambda_3$.

These abelian gauge multiplets are not inevitably light; they may
acquire a high mass from fluxes or be projected away by orientifold
planes.  However, given the tremendous multiplicity of independent
cycles on a realistic compactification manifold, it is not
unreasonable to expect a variety of massless $U(1)$ fields to survive
to low energies. Indeed, the only objects charged under RR forms are
non-perturbative D-brane states, and so the 4d particle states charged
under RR $U(1)$s arise from D-branes wrapping the corresponding
cycles; there are typically no light charged states that could higgs
the $U(1)$s. 

Without light charged states, at low energies the RR $U(1)$s interact
with the Standard Model fields either through higher-dimensional
operators or through renormalizable kinetic mixing with the
hypercharge group $U(1)_Y$. Amusingly, the absence of light charged
states for RR $U(1)$s naturally mitigates limits on such kinetic mixing
from astrophysics and laboratory searches for millicharged particles
\cite{Davidson:1993sj,
  Davidson:2000hf,Dubovsky:2003yn,Melchiorri:2007sq}. Without light charged states, the
kinetic mixing between hypercharge and RR photons can be removed by
field redefinition without introducing any physical effects, apart
from a shift in the hypercharge gauge coupling.  However, the
situation becomes significantly more interesting in the presence of
low energy supersymmetry breaking. In this case massless RR photons
are accompanied by their light fermionic superpartners, the so-called
photini. Unlike vectors, which are protected by gauge invariance, RR
photini acquire masses of order the gravitino mass $m_{3/2}$ as a
result of SUSY breaking.  If the dominant source of SUSY breaking for
the MSSM also comes from gravity mediation, then these photini masses
are of the same order as the MSSM soft masses.

As a consequence of a non-trivial photini mass matrix, the mixing of
RR photini with the bino cannot be rotated away by the same field
redefinition that decouples the vector fields, and thus has observable
effects.  For the purposes of LHC phenomenology, the significant
result of this mixing is the extension of the MSSM neutralino sector
by one or more new states mixed with the bino through the gaugino mass
matrix.  This leads to a variety of possible signatures depending on
the amount of mixing and the size of mass splittings between photini,
including extended supersymmetric cascades, displaced vertices, and --
in the case of a charged LOSP that stops in the detector --
out-of-time decays of the LOSP to photini.  While these signatures may
be spectacular, there is a fairly pedestrian possibility that may be
of greater relevance to current LHC limits: light and moderately-mixed
photini. In this case, supersymmetric cascades typically end in the
LOSP, followed by a prompt and non-displaced decay to one or two of
the lightest photini, which then escape the detector. These final
decays may substantially erode the $\slashed{E}_T$ present in the
event.

\subsection{Photini in the IR}

Setting aside the UV motivation, we may focus on the IR effective theory of additional $U(1)$ gauge multiplets kinetically mixed with the MSSM. In a supersymmetric theory, kinetic mixing between $U(1)$s takes the form \cite{Dienes:1996zr}
\beq
  \mathcal{L}_{\rm gauge}~=~ \frac{1}{32}\int d^2\theta \,\left (W_aW_a+W_bW_b-2\epsilon W_aW_b,
\right )
\eeq
where $W_a$ and $W_b$ are the chiral gauge field strength superfields for
the two gauge symmetries (e.g.~$W_a =\bar D^2D V_a$ for the $U(1)_a$ vector superfield $V_a$). We may bring the pure gauge portion of the Lagrangian to canonical form via the shift $W_b \to W'_b=W_b- \epsilon W_a$. 
This renders the gauge Lagrangian diagonal,
\beq
\mathcal{L}_{\rm gauge}~=~\frac{1}{32}\int d^2\theta \,\left (W_a W_a+ W_b' W_b' \right )~,
\eeq
and shifts the visible-sector gauge coupling by an amount
\beq
g_a \rightarrow g_a / \sqrt{1 - \epsilon^2}~.
\label{couplingshift}
\eeq
Note that if there are many photini with large mixings, this may spoil the successful prediction of gauge coupling unification based on measurements at the weak scale.

If there are no light states charged under $U(1)_b,$ then this field redefinition has no effect on the interactions of states charged under $U(1)_a.$ The hidden sector photon decouples entirely, leaving only a shift in the hypercharge gauge coupling \cite{Holdom:1990xp}. However, the story changes when the remainder of the supermultiplet is considered. Although the $U(1)_b$ gauge boson may be decoupled by field redefinitions, the gaugino $\lambda_b$ may still mix with the visible sector via off-diagonal terms in the gaugino mass matrix \cite{Abel:2008ai, Ibarra:2008kn}. If this mass matrix is not diagonalized by the same field redefinition, the hidden-sector gaugino retains interactions with the visible sector even in the absence of light states charged directly under $U(1)_b$.

Focusing on the gauge fermions, after supersymmetry is broken the mixings between the photini and MSSM gauginos are encoded in the Lagrangian terms 
\beq
\delta \mathcal{L} \supset i Z_{IJ} \lambda_I^\dag \slash \partial \lambda_J - m_{IJ}  \lambda_I  \lambda_J
\eeq
where $I,J$ run across the bino $\tilde B$ and $n$ photini $\tilde \gamma_i;$ the $Z_{IJ}$ encode arbitrary kinetic mixing, while the $m_{IJ}$ are soft masses generated by supersymmetry breaking. As with the gauge kinetic terms, the gaugino kinetic terms may be diagonalized via field redefinitions so that $Z_{IJ} \rightarrow \delta_{IJ}$ and $m_{IJ} \rightarrow m_{IJ}'$. If the kinetic terms are made canonical by the transformation $\lambda_I \rightarrow \lambda'_I = P_{IJ}^{-1} \lambda_J,$ then we have $m_{IJ}' = P^{\dag}_{IK} m_{KL} P_{LJ}$ and mixing persists in the mass matrix. Moreover, since the final physical mixing among the gauginos depends on the mass
matrix mixing, the gauge-coupling unification constraint on the amount of kinetic mixing with hypercharge does not limit the size of the mixing among gauginos.

To study the neutralino mass eigenstates, we diagonalize the gaugino mass matrix $\mathbf{m}$ via $\mathbf{m}_{D} = \mathbf{U}^* \mathbf{m} \, \mathbf{U}^{-1},$ where $\mathbf{U}$ is a unitary matrix. The mass eigenstate neutralinos $\tilde N_I$ may then be written as 
\beq
\tilde N_I = U_{IJ} \lambda_J
\eeq
where $I, J = 1, ..., n+4$ runs over the four MSSM neutralinos and the $n$ photini; $U_{IJ}$ are the components of the matrix $\mathbf{U},$ and $\lambda_I = (\tilde B, \tilde W, \tilde H_d, \tilde H_u, \tilde \gamma_1, ..., \tilde \gamma_n)$ are the gauge eigenstate gauginos with canonical kinetic terms. As long as the mixing is not $\mathcal{O}(1)$, the neutralinos decompose into mostly-MSSM and mostly-photino states. Consequently, we may think of the $\tilde N_a \, (a = 1, ..., 4)$ as mostly-MSSM neutralinos, and the $\tilde N_i \, (i = 5, ...., n+4)$ as mostly-photino neutralinos. For the most part, we need not concern ourselves with the majority of the photini. If mixing is large, the effects on collider phenomenology are dominated by one or two photini at the bottom of the spectrum; transitions through heavier photini are kinematically suppressed.

In this work we restrict our attention to the case where mixing is
relatively large and erosion of $\slashed{E}_T$ is maximized:
$\epsilon_i \gtrsim 0.1$. Although our interest lies in collider
physics, there are several noteworthy consequences for
phenomenology. Mixing of this order, while large, is consistent with
gauge coupling unification, since it only results in a percent-level
shift in the hypercharge gauge coupling. As long as there are no more
than a few such strongly-mixed photini, this correction to gauge
coupling unification is smaller than typical threshold
corrections. And as noted above, the dominant contribution to mixing
may not even come from kinetic terms, in which case even this
constraint is mitigated. 

Large mixing also has favorable consequences
for cosmology if the strongly-mixed photini are the lightest in the
spectrum. Generic photini with small mixings make poor dark matter
candidates, since their thermal relic abundance scales as
$\epsilon^{-4}$ times the equivalent bino abundance, leading to a
substantial overdensity of dark matter (though this improves to an
$\epsilon^{-2}$ scaling if the photino LSP is close enough in mass to
the higgsino for efficient coannihilation). For weakly-mixed
photini this requires an alternative cosmology or non-thermal origin
for the dark matter relic abundance. But if the LSP photino mixing
parameter is $ \gtrsim 0.1$ the thermal relic abundance may itself be
adequate, particularly if coannihilation is effective.\footnote{We note that coannihilation is unavailable in the detailed spectra considered in this paper due to the lightness of the photini, but simply wish to emphasize the favorable cosmology of this parametric limit.}  In this
respect, a few light photini with large mixing are favorable for more
than just LHC phenomenology.

\subsection{Photini at the LHC}

The addition of light photini to the supersymmetric spectrum can
effectively hide SUSY at the LHC by reducing the $\slashed{E}_T$ of events with supersymmetric particles. In the presence of
photini, the lightest particle in the MSSM can decay further to these
light states through $Z$ or higgs boson emission.  If the photini are
light and have large mixing with the bino (i.e.~the bino decay is
prompt), then the bosons will carry away most of the energy of the
bino as SM decay products, dramatically reducing the missing
transverse energy in an event and thus the efficiencies of
experimental searches.  To see this explicitly, consider the two-body
decay of a bino into a $Z$ and a light relativistic photino (in the rest frame of the bino). Then the
fraction of the bino energy carried away by the photino is
\begin{equation}
 \frac{E_{\tilde{N}}}{E_1} = \frac{m_1^2-m_Z^2}{2m_1^2} + \mathcal{O}\left(\frac{m_{\tilde{N}}^2}{m_Z^2}\right), \label{eq:Ereduction}
\end{equation}
where $m_1 - m_{\tilde{N}} > m_Z \gg m_{\tilde{N}}$ and $m_1, m_Z,$
and $m_{\tilde{N}}$ are the bino, $Z$ and photino masses,
respectively. The fraction tends to zero when the bino mass approaches
that of the $Z$ boson.  To get the total $\slashed{E}_T$ in a typical
SUSY event at the LHC, one has to sum up the momenta of the photini
and any neutrinos from $Z$ decays on both sides of the SUSY decay
chain, and take the transverse component of this four-momentum sum.
Hence the missing energy reduction of \eqref{eq:Ereduction} degrades
somewhat, but roughly persists when the binos are only moderately
boosted and have momenta oriented in uncorrelated directions.  If the
bino decay occurs dominantly through light higgs (rather than $Z$) emission,
there is no irreducible $\slashed{E}_T$ from the $\sim20\%$ branching
ratio of $Z$'s to neutrinos, reducing the total $\slashed{E}_T$ even
further (though of course there is still a fraction of $\slashed{E}_T$ in
higgs decay products as well).

Since most SUSY searches at particle accelerators look for events
with a large amount of $\slashed{E}_T$, this extra step in
the decay cascade tends to diminish the SUSY signal in the canonical
ATLAS and CMS searches. We show that this effect significantly weakens
the exclusion limits on SUSY from the most constraining hadronic
searches at ATLAS and CMS in the next section.

For concreteness, we consider a SUSY spectrum with a mostly-bino LOSP (the mass eigenstate $\tilde{N}_1$). The
mixings between photini and the bino, inherited from the kinetic
mixings of the massless $U(1)$ fields, lead to a main decay channel
through $Z$-boson emission: $\tilde{N}_1 \rightarrow \tilde{N}_i +
f\overline{f}$. If the $Z$ is produced off-shell, the decay rate is of
the order
\begin{align}
  \label{eq:1}
  \Gamma(\tilde{N}_1\rightarrow \tilde{N}_i + f\overline{f}) \simeq
  \frac{1}{192\pi^2} \frac{\alpha_{W}}{c_{W}^2} \eta^4
\,\epsilon_i^2\,\frac{(\delta
    m_i)^5}{m_Z^4} \, \mathrm{Br}(Z\rightarrow f\overline{f}),
\end{align}
where $\delta m_i = |m_1 - m_i|$ is the mass splitting between the bino and the
$i$-th neutralino and the parameter $\eta~\sim \mathcal{O}(m_Z/M_{1,2}) \sim \mathcal{O}(m_Z/\mu) $ characterizes
the size of the mixings of the bino and higgsinos in the MSSM. In the
spectra we consider here, $\eta \gtrsim 0.1$. For a generic case where
mixing parameters $\epsilon_i$ of the photini are of the same order,
we can see from \eqref{eq:1} that the decay to the lightest photino
will in general dominate. In the case of gravity mediation, photini
masses are of the same order as soft masses in the MSSM, and for 
topologically complex manifold compactifications which produce a
multitude of photini, generically one or more of the lightest
neutralino mass states will have a mass below the bino-$Z$ mass
splitting, $|m_i| < m_1 - m_Z$.\footnote{In gauge mediation the photini masses may be significantly lighter, of order the gravitino mass. In this case the phenomenology changes qualitatively due to the presence of the gravitino at the end of SUSY decay chains, and is likely to resemble that of goldstini \cite{Cheung:2010mc,Cheung:2010qf, Craig:2010yf}. While potentially very interesting, we will not consider the gauge-mediated scenario in detail here. }

In this case, the $Z$-boson is on-shell and the decay rate is given by \cite{Gunion:1987yh}
\begin{align}
  \label{eq:2}
   \Gamma(\tilde{N}_1\rightarrow \tilde{N}_J +Z) \simeq
    \frac{\alpha_{W}}{2 c_{W}^2} \eta^4
\,\epsilon_i^2\,\frac{(m_1-m_Z)^2}{m_1} 
\end{align}
for $m_i \ll m_1, m_Z$.
Another possibility is for the decay to go through the lightest
neutral higgs,
\begin{align}
  \label{eq:3}
  \Gamma(\tilde{N}_1\rightarrow \tilde{N}_J + f\overline{f}) \simeq
\frac{ \alpha_{W} }{2} \eta^2
\,\epsilon_i^2\,\frac{(
    m_1-m_H)^2}{m_1}
\end{align}
again for  $m_i \ll m_1, m_H.$
The relative ratio between higgs decays and $Z$ decays depends on the
mixing in the neutralino matrix and the relevant mass splittings.\footnote{Here we assume $h$ is the lightest scalar in the higgs sector. Of course, as the mass of the LOSP increases, it is also possible for decays to involve the heavier higgs states $A, H,$ and $H^\pm$. For simplicity, we take these states to be decoupled, though including them does not radically change the phenomenology.}

Other subleading decays for the LOSP are through intermediate squarks
and sleptons, or loop-suppressed photon emission; these processes
produce similar reduction in SUSY limits but have different discovery
channels. For the parameters considered in this work, these ancillary processes are unimportant.
In particular, although there may be two-body decays of the
LOSP via a photon, the couplings of photini to the photon are
suppressed by an additional loop factor. Since (for the spectra
considered here) decays involving photons compete with open two-body
decays to the $Z$ or higgs, they are substantially suppressed and
typically do not provide interesting limits from e.g. $n\gamma +
\slashed{E}_T$ searches. In this sense the phenomenology of the MSSM
with light photini is substantially different from that of gauge
mediation, where two-body LOSP decays involving photons often dominate.

\begin{figure} [t]
  \begin{center}
    \subfigure[]{\includegraphics[trim = 5mm 2mm 26mm 10mm, clip, width
      = 0.33\textwidth]{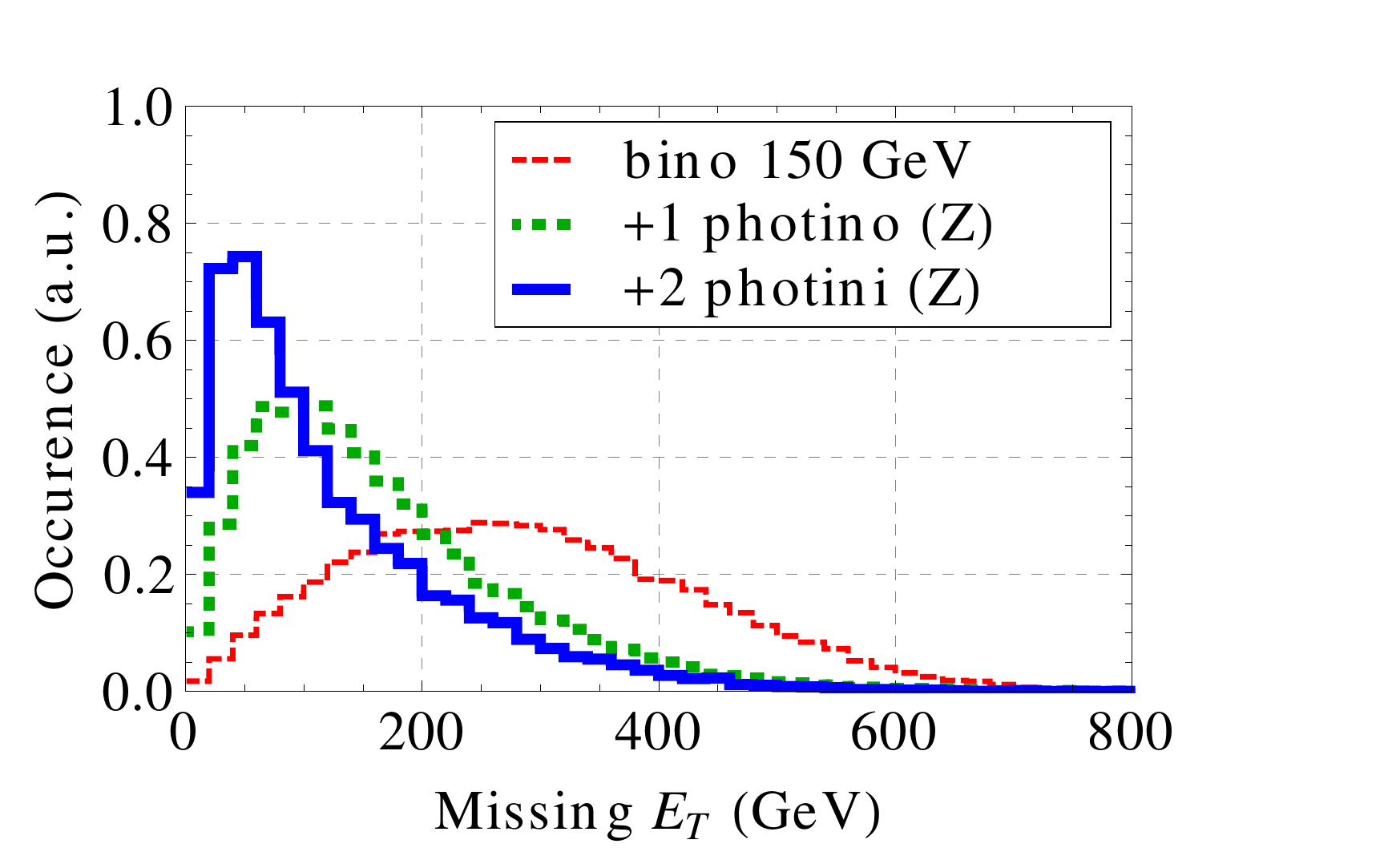}
   \label{fig:met_number}}
\hspace{-4mm}
 \subfigure[]{\includegraphics[trim =  5mm 2mm 26mm 10mm, clip, width
      = 0.33\textwidth]{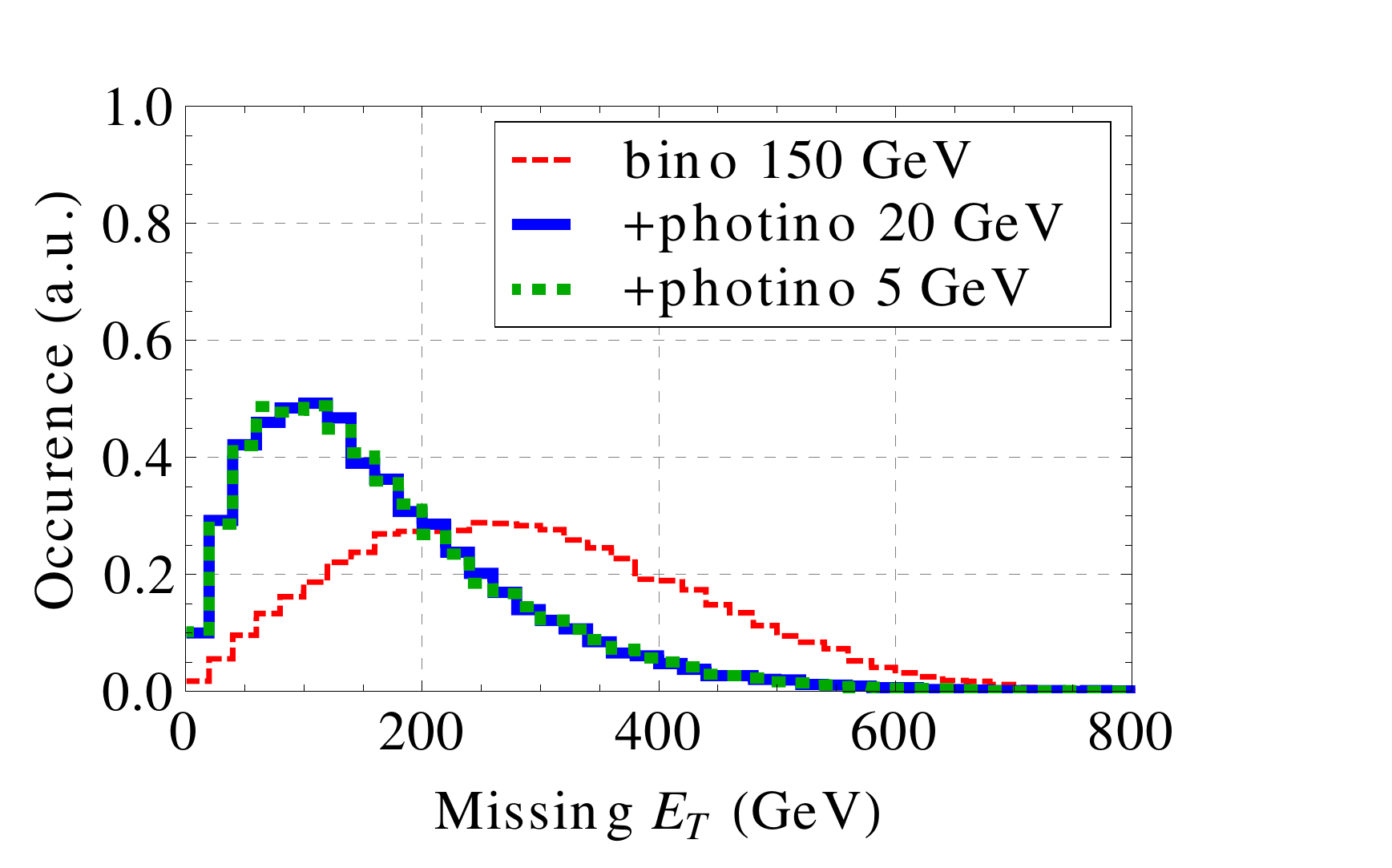}
   \label{fig:met_mass}}
\hspace{-4mm}
    \subfigure[]{\includegraphics[trim = 5mm 2mm 26mm 10mm, clip, width
      = 0.33\textwidth]{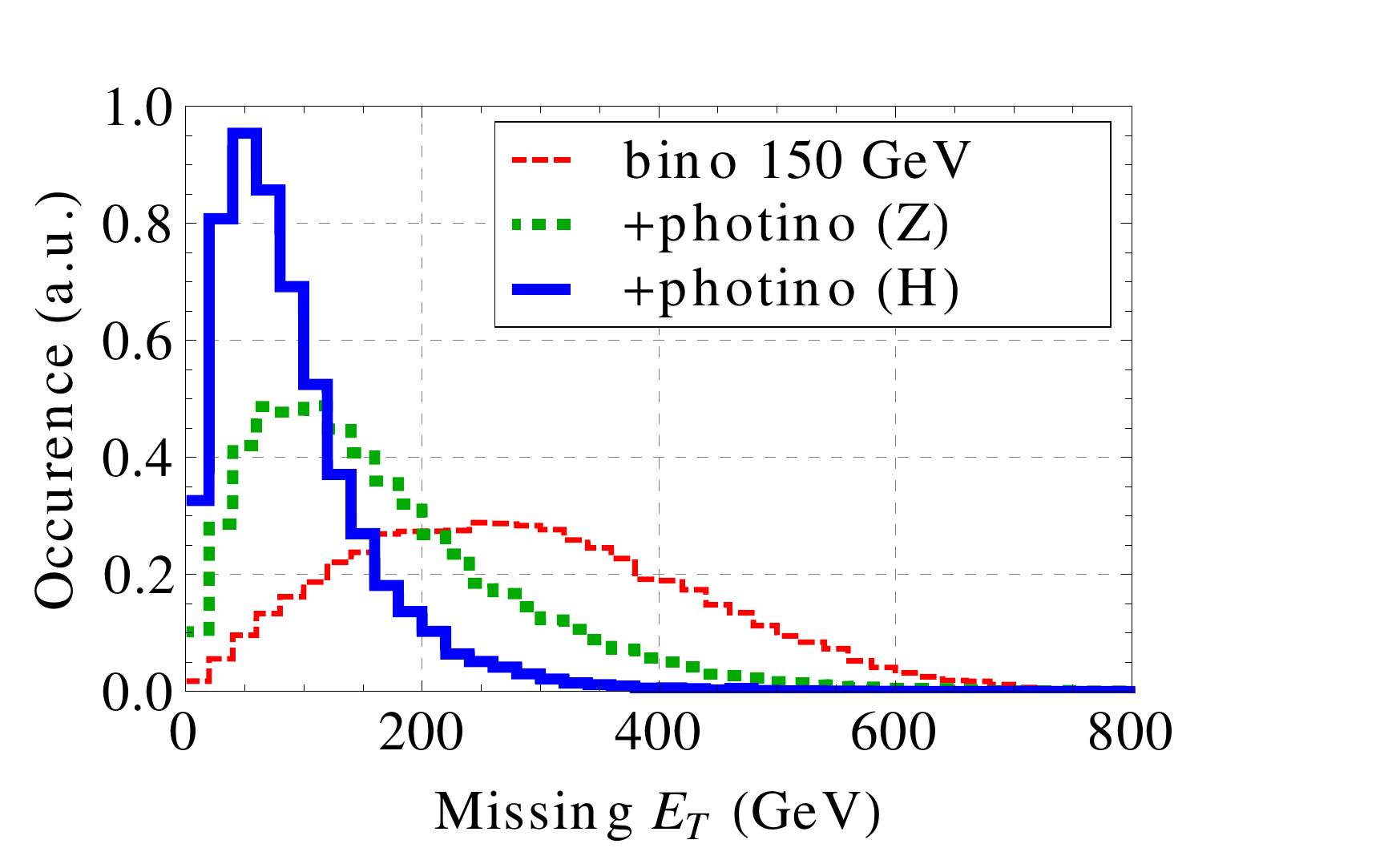}
    \label{fig:met_higgs}}
  \end{center}
\label{fig:met}
\vspace{-4mm}
\caption[Optional caption for list of figures]{\small{$\slashed{E}_T$ histograms for different photini spectra. We
    present the effect of \subref{fig:met_number} number of photini in
    the decay chain: no photini, one $5\gev$ photino or cascade
    through $20\gev$ and $5\gev$ photini; \subref{fig:met_mass} mass
    of the lightest photino; \subref{fig:met_higgs} the SM particle in
    the cascade: bino to $Z$ + photino or bino to 125 GeV higgs + photino. The SUSY model is a simplified spectrum with
    a 750 GeV gluino and 150 GeV bino LOSP, with the rest of the
    sparticles decoupled; this benchmark point is currently excluded by experiment
    as a standard scenario but allowed in the presence of photini}.
    }
\end{figure}

If the bino undergoes a cascade decay through several photini, missing
energy becomes converted into $Z$ or higgs decay products at every
step of the cascade. In Figure~\ref{fig:met_number} we present the
missing energy distribution for the case that the bino-like
$\tilde{N}_1$ is the true LSP versus the LOSP that can decay through a
chain of $1$ or $2$ photini via $Z$ bosons. We consider a sample
simplified spectrum with $m_{\tilde{g}} =750\gev, \,m_{\tilde{N}_1} =
150\gev$ and photini at 20 GeV and 5 GeV. The second $Z$ is far
off-shell in the 2-photini cascade, so the reduction in missing energy
between steps $1$ and $2$ is relatively smaller. Adding more steps to
the cascade will further decrease the missing energy and hide the
signal, but for a generic selection of kinetic mixings between
photini, the decay to the lightest state or lightest two states will
generally dominate.  Therefore, we only consider decays to one or two
relatively light photini with masses $5\gev$ and $20\gev$.

Missing energy can be dissipated in a decay chain of several photini;
usually even one photino is sufficient to dramatically reduce the
missing energy in an event. In Figure~\ref{fig:met_mass} we present
the missing energy spectrum in the same model, with the $\tilde{N}_1$
decaying through a $Z$ in one step to a photino of mass $20\gev$ or
$5\gev$; since the $Z$ is on-shell in both cases and the photini are
much lighter than the bino, there is little difference between the two
cases. In the remainder of the paper we will consider a single $5\gev$
photino model, which generalizes to other masses as is clear from
Figure~\ref{fig:met_mass}. In Figure~\ref{fig:met_higgs}, we see that
higgs emission is even more efficient at reducing $\slashed{E}_T$ than
$Z$ emission; this is due to the fact that the $Z$ decays invisibly to
neutrinos $20\%$ of the time, while the higgs is heavier thus carrying
away a larger fraction of the bino's energy and has a much smaller
fully invisible branching fraction.

\section{Hiding SUSY} 

In this section, we explore the phenomenology of photini in more detail in
the context of current ATLAS and CMS hadronic searches. In
Section~\ref{sec:simdetails} we describe the example SUSY spectra we
use for our analysis and the implementation of our Monte Carlo-based
simulation, while in Section~\ref{sec:hadronicsearches} we probe
the change in exclusion limits of current hadronic searches for SUSY
at the LHC in the presence of photini. Although presented in the context of photini, our results may be generalized to any theory in which one or more additional light fermions mix moderately with the bino, such that SUSY cascades proceed through a bino to the LSP via emission of on-shell $Z$ or higgs bosons.

\subsection{Simulation details}\label{sec:simdetails}

In order to accurately analyze how photini affect experimental limits
on supersymmetry, we consider the leading experimental searches with
jets and
$\slashed{E}_T$ \cite{ref:cmsjetsmet1fb,ref:atlas2to6jets5fb,ref:alphaT1fb,ref:razor5fb}
and compare models with a standard neutralino LSP and with
photini. The mechanism by which photini alter LHC limits is general for any SUSY model with a
neutralino LOSP; in this paper, we focus on three frequently-presented
parameter sets which allow direct comparison with existing
experimental exclusion limits:
\begin{enumerate}
\item a simplified model with a gluino, first- and
  second-generation squarks and a light bino LOSP, with the remainder of the sparticles
decoupled at $3.5\tev$ \cite{Alves:2011wf,Alwall:2008ag}.
\item a simplified
model with a gluino and bino LOSP with the remainder of the sparticles
decoupled at $3.5\tev$ \cite{Alves:2011wf,Alwall:2008ag}.
\item the constrained MSSM (cMSSM) model \cite{Kane:1993td, Chamseddine:1982jx, Barbieri:1982eh, Ibanez:1982ee, Hall:1983iz, Ohta:1982wn} with a mostly-bino
LOSP, with parameter choices $A_0 = 0$, $\tan \beta = 10$, and $\mu >
0$ for varying values of $m_0$ and $m_{1/2}$. The low energy spectra are
generated with \texttt{SOFTSUSY 2.0.18} \cite{Allanach:2001kg}. % and $m_t = 173.2\gev$)
\end{enumerate}
% For the gluino-squark-bino simplified model, we decouple the sleptons and third-generations squarks to $3.5\tev$, set the bino mass to 195 GeV, and scan over the masses of the gluino and light-flavored squarks.  For the gluino-bino simplified model, we assume sfermion masses of $3.5\tev$ and implement a scan over gluino and LOSP masses. For the cMSSM spectra,
% we use the parameters, and
% scan over the remaining parameters $m_0,\,m_{1/2}$ in the range
% $500\gev~\leq~m_0~\leq~1000\gev$ and
% $250\gev~\leq~m_{1/2}~\leq~500\gev$; 

%\begin{figure}[t]
%\begin{center}
% \includegraphics[trim = 0mm 0mm 0mm 0mm, clip, width
 %     = 0.4\textwidth]{fig/T1.pdf}
  % \caption{\small{Simplified model (T1) kinematics. The
   %    gluinos decay exclusively to light flavor jets [cite].}}  
%\label{fig:T1}
%\end{center}
%\end{figure}

We use the superspace module of \texttt{Feynrules v1.4.0}
\cite{Christensen:2008py} (altered to include the dynamics of extra
abelian superfields without light charged matter under them) to incorporate the photini $\tilde{N}_i$ in
\texttt{UFO} model files for supersymmetric spectra \cite{Degrande:2011ua}. We add $n$ photini
of mass $m_i$, each with kinetic mixing $\epsilon_i=\mathcal{O}(0.1)$
with the bino and diagonalize the $(4+n)\times(4+n)$ neutralino mass
matrix at tree level to find the neutralino mass eigenstates.

For the hard SUSY processes, we simulate LHC events at $\sqrt{s} =
7\tev$ for the leading squark-squark, squark-gluino and gluino-gluino
production channels using \texttt{MadGraph5 v1.4.03} \cite{Alwall2011}, 
compute NLO cross-sections for the processes in \texttt{Prospino v2.1}
\cite{Beenakker:1996ed}, and scan over the two-dimensional parameter
spaces in the three types of spectra above. We differentiate
between the three cases in implementing cascade
decays of the squarks and gluinos in \texttt{Pythia v6.426}
\cite{Sjostrand:2006za} to produce parton-level event samples with the
bino as the LSP (``bino samples''). We use the bino samples to
validate our analysis and ensure that the limits we set on photini
spectra are consistent with experimental limits up to acceptable simulation uncertainties.

In the case of squark-gluino-bino and gluino-bino simplified models,
we simulate the ``photini samples'' by adding an extra step in the
cascade with the LOSP (the LSP in the ``bino samples'') decaying via $Z$ or higgs emission, with a higgs
mass of $125\gev$ as suggested by recent experiments
\cite{ATLAS2012higgs,CMS2012higgs}. To generate the photini samples in
the cMSSM, we run \texttt{Pythia} directly on the cMSSM bino samples
with bino decays to photini through $Z$ and higgs bosons; we calculate
the $Z$ versus higgs boson emission rates in \texttt{Madgraph} with
photini \texttt{UFO} model file using a higgs mass obtained from
\texttt{SOFTSUSY}.\footnote{Note that this fixes the higgs mass to be
  somewhat lighter than the recent tantalizing hints of the particle
  at a mass of $\sim125\gev$.  However, this small change would not
  significantly alter our results.}

Finally, in order to account for basic detector effects, we use the simplified
detector simulator \texttt{PGS v120404} \cite{conwaypgs} (as part of
the \texttt{Pythia-PGS v2.1.16} implementation in \texttt{Madgraph5})
with its default LHC settings and lepton isolation.

\subsection{Sensitivity of hadronic searches}\label{sec:hadronicsearches}

\begin{figure} [t]
  \begin{center}
\includegraphics[trim = 0mm 0mm 0mm 0mm, clip, width
      = 0.55\textwidth]{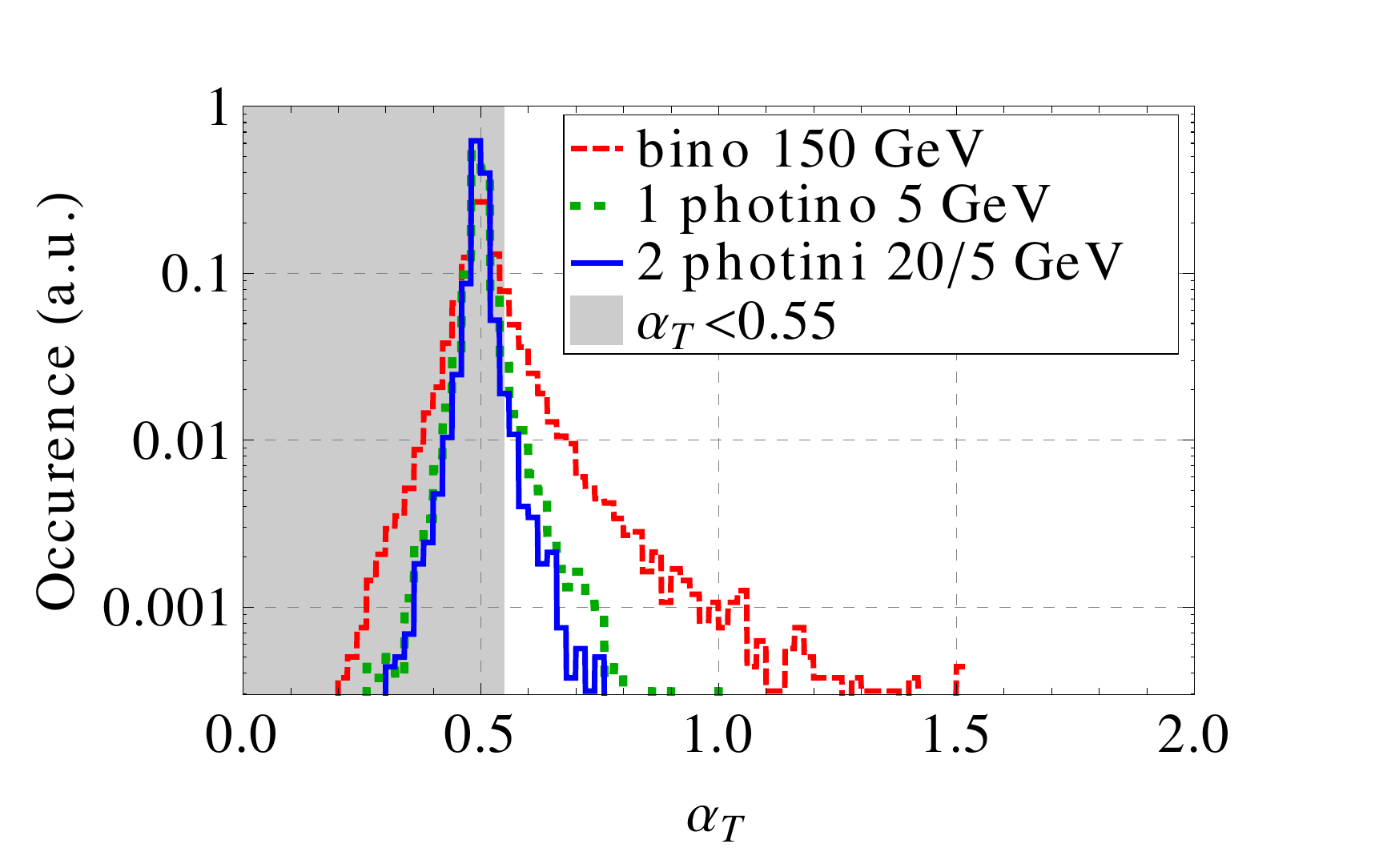}
\vspace{-4mm}
  \end{center}
  \caption[]{\small{The histogram of the experimental $\alpha_T$
      variable for the simplified model of $750\gev$ gluino and a
      $150\gev$ bino; the signal region is $\alpha_T \geq 0.55$ \cite{ref:alphaT1fb}. The
      three curves show the $\alpha_T$ distribution for spectra in
      which the bino is stable (red, dashed), the bino decays to a $5\gev$
      photino by $Z$ emission (green, dotted), and the bino dominantly
      cascades through two photini of $20\gev$ and $5\gev$ by double
      $Z$ emission (blue, solid). }}
\label{fig:alphaT}
\end{figure}

\begin{figure} [t]
  \begin{center}
    \subfigure[]{\includegraphics[trim = 0mm 0mm 0mm 0mm, clip, width
      = 0.32\textwidth]{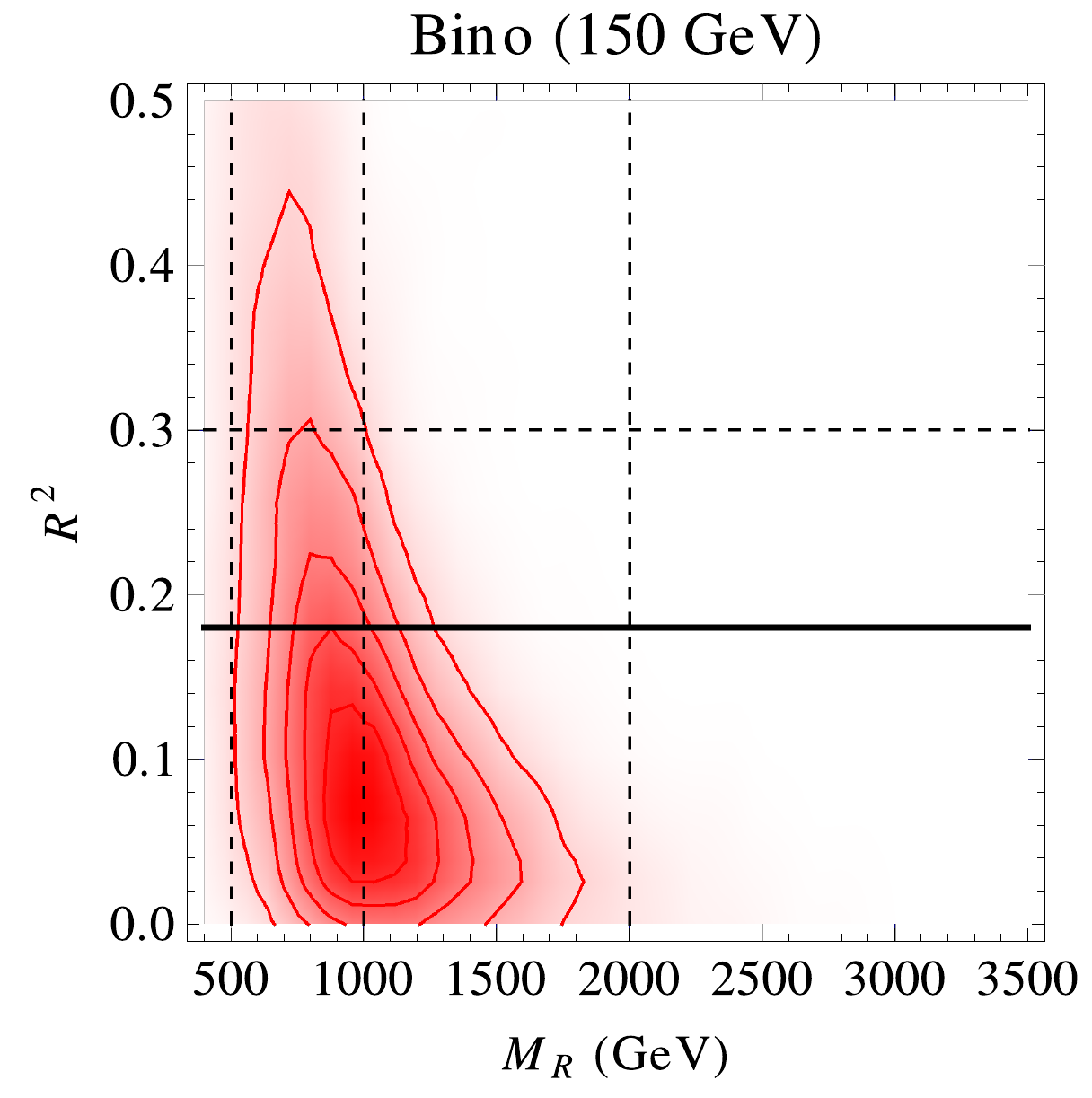}}
\hspace{-4mm}
    \subfigure[]{\includegraphics[trim = 0mm 0mm 0mm 0mm, clip, width
      = 0.32\textwidth]{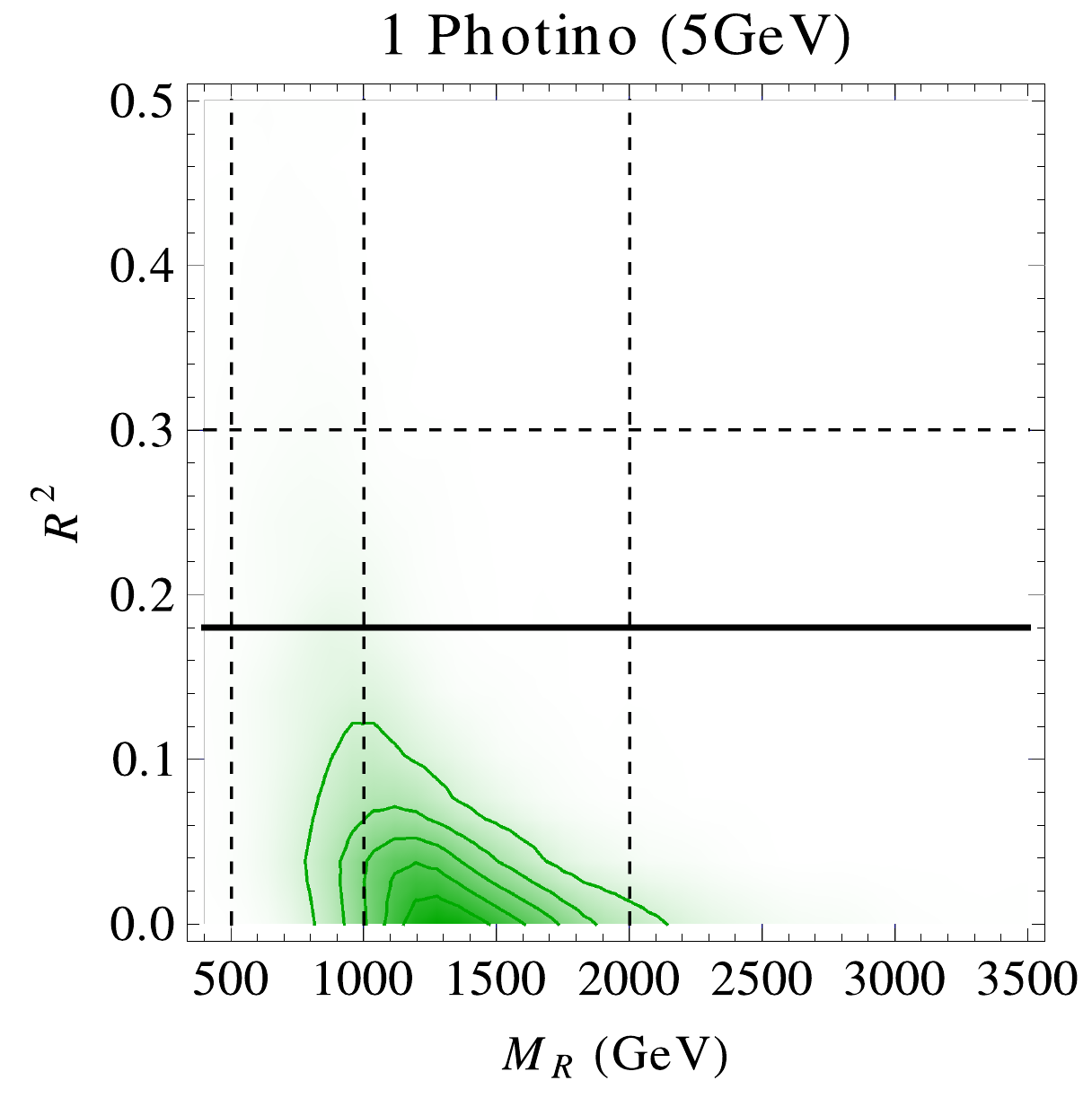}}
\hspace{-4mm}
    \subfigure[]{\includegraphics[trim = 0mm 0mm 0mm 0mm, clip, width
      = 0.32\textwidth]{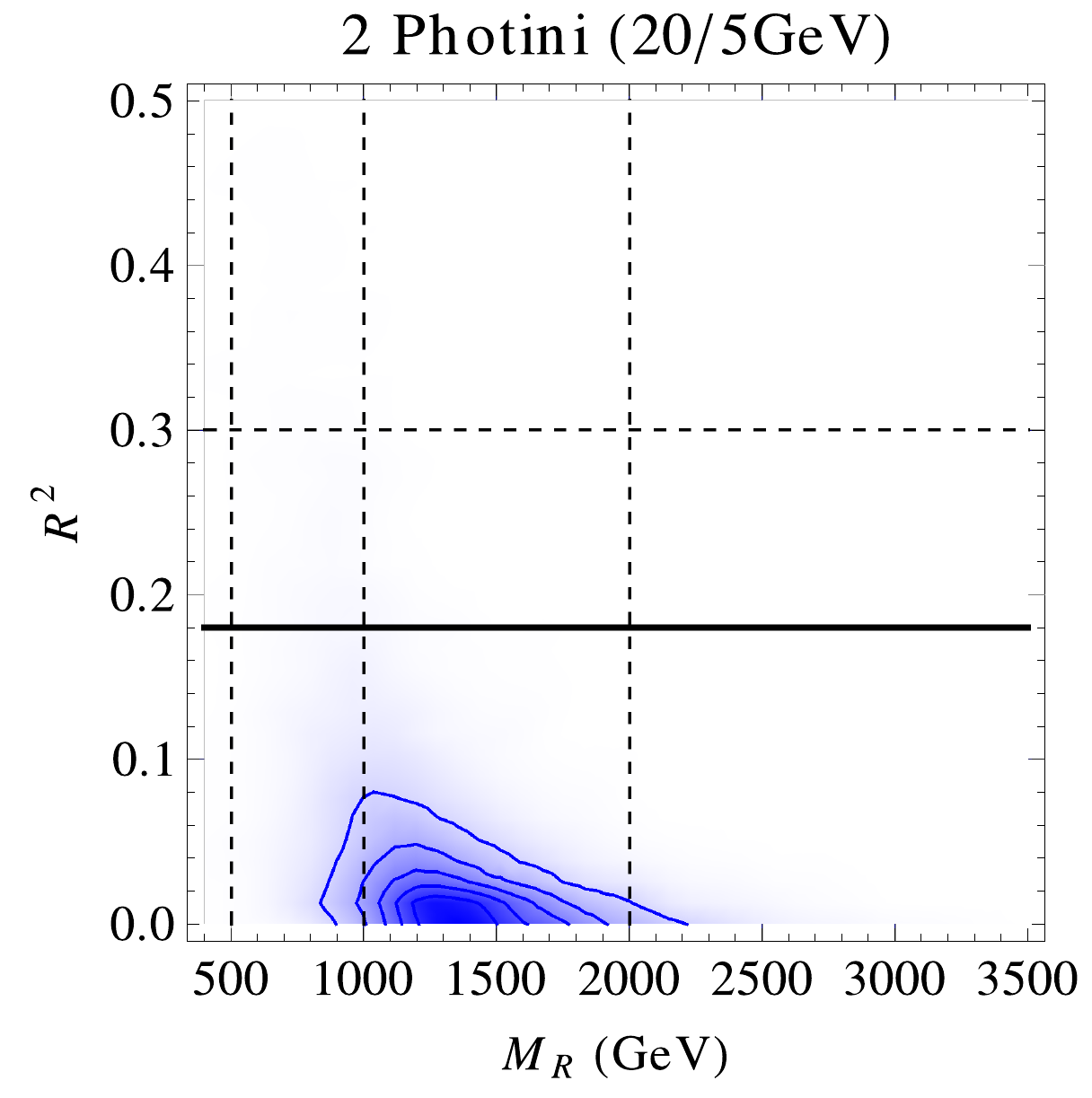}}
 \end{center}
\vspace{-4mm}
\caption[]{\small{Razor variables density plot excluding all other
    selection cuts for the simplified model of $750\gev$ gluino and a
    $150\gev$ bino.  The three plots concern spectra in which (a) the
    bino is stable; (b) the bino decays to a $5\gev$ photino by $Z$
    emission; (c) the bino dominantly cascades through two photini of
    $20\gev$ and $5\gev$ by double $Z$ emission. The six signal
    regions from the CMS analysis \cite{ref:razor5fb} correspond to the
    six regions outlined by the grid lines above the bold black line
    (i.e. $R^2 > 0.18$ and $M_R>500\gev$); the SUSY signal in these
    regions is decimated if the bino can decay to one or more
    photini.}}
\label{fig:razor}
\end{figure}
In order to evaluate the impact of the missing energy spectrum shifts
demonstrated above, we perform a detailed analysis of experimental
limits and how search efficiencies change with photini added to the
spectrum. We focus on LHC searches; for the range of sparticle masses
we consider ($m_{1/2} >250\gev$ or $m_{\tilde{g}}>400$), Tevatron and
LEP searches provide no relevant limits.

First, we consider additional observables that have been defined by
the CMS collaboration in order to enhance signal sensitivity, in
particular $\alpha_T$ (Figure~\ref{fig:alphaT}) and razor variables
(Figure~\ref{fig:razor}) \cite{ref:alphaT1fb, ref:razor5fb}. Even
though these analyses are carefully optimized to search for SUSY
signals, they still fundamentally rely on missing energy as a handle
on the event, and are thus less effective in the presence of photini.
Both $\alpha_T$ and the razor variable $R$ are defined as ratios of
energies in the event in
order to maximize the signal to background ratio of SUSY versus
Standard Model events  (see Appendix~\ref{app:cuts} for details).  We see that adding one photino to the decay
chain makes the events more Standard-Model-like by diffusing the
missing energy in the event (red versus green curves in
Figures~\ref{fig:alphaT}, \ref{fig:razor}). Adding a second photino to
the cascade reduces the efficiency in the signal regions further.

To quantify the impact of signal reduction described above, we impose
selection criteria that mimic those of the experimental LHC searches
on the Monte Carlo samples.  For simplicity, we include only the
leading selection cuts for the experimental searches.  In
Appendix~\ref{app:cuts}, we present our exact event selection criteria
for each of the relevant hadronic searches on our PGS samples, which
already account for some detector efficiencies and lepton isolation
cuts.

To set exclusion limits, we use the published expected and observed
event counts in each experimental search region. We  convolve the Poisson statistical errors with
Gaussian systematics using \cite{limitcalc}; for searches ($\alpha_T$) without published
errors, we estimate the systematic error to be $30\%$. For the
simplified model search we also use a varying error on the efficiency
in the gluino-bino plane as published in the $36$ pb$^{-1}$ results
\cite{ref:alphaT36pb}. Despite our course-grained analysis, we find that
our Monte Carlo sensitivities for cMSSM and simplified spectra (with a bino as the
LSP) match well with those of CMS and ATLAS (Figures \ref{fig:simplifiedexclusion}
and \ref{fig:cmssmexclusion}).

\begin{figure}[t]
\begin{center}
    \subfigure[]{\includegraphics[trim = 0mm 0mm 0mm 0mm, clip, width
      = 0.48\textwidth]{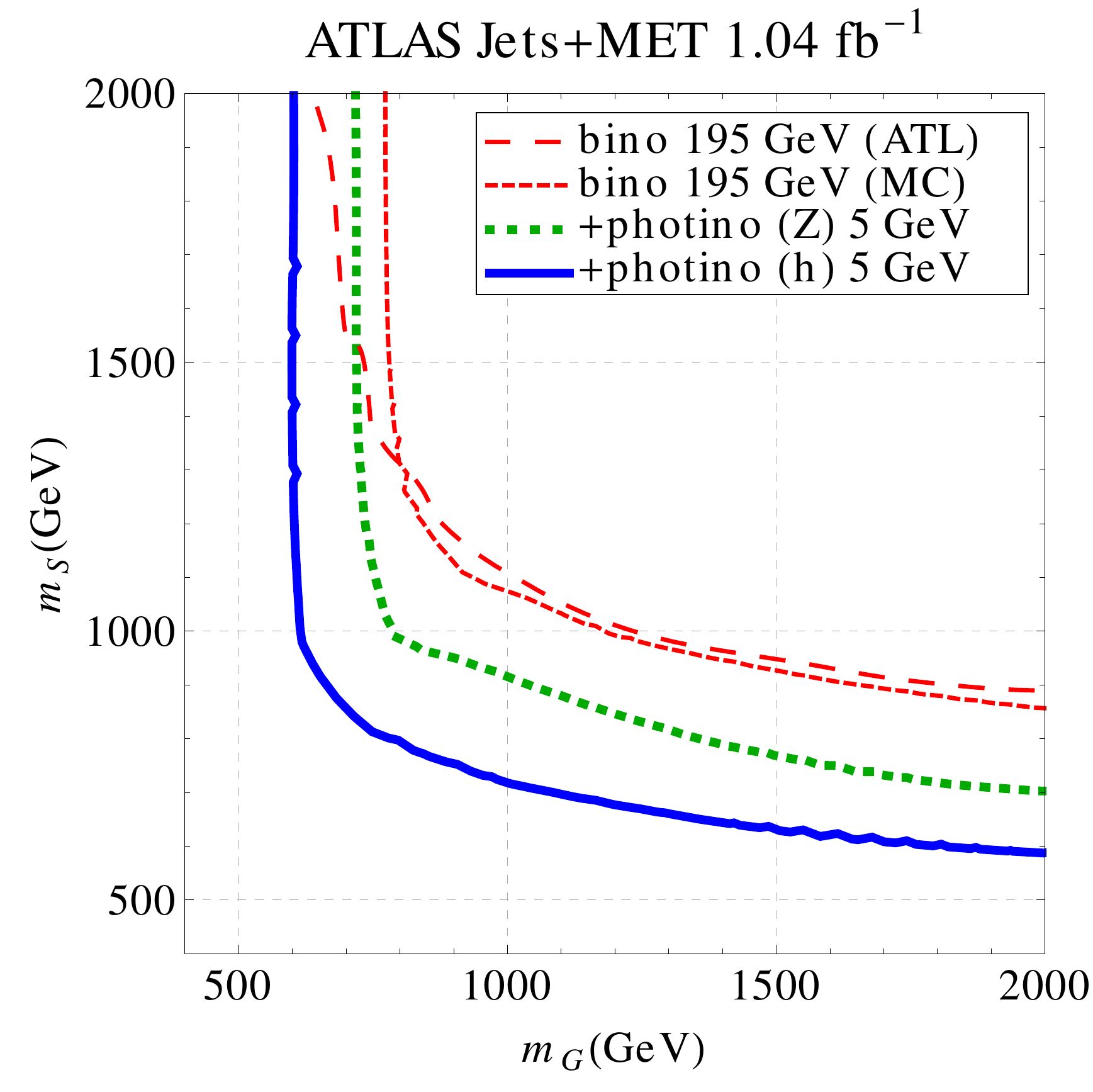}\label{fig:atlasexclusiona}}
\hspace{-2mm}
    \subfigure[]{\includegraphics[trim = 0mm 0mm 0mm 0mm, clip, width
      = 0.48\textwidth]{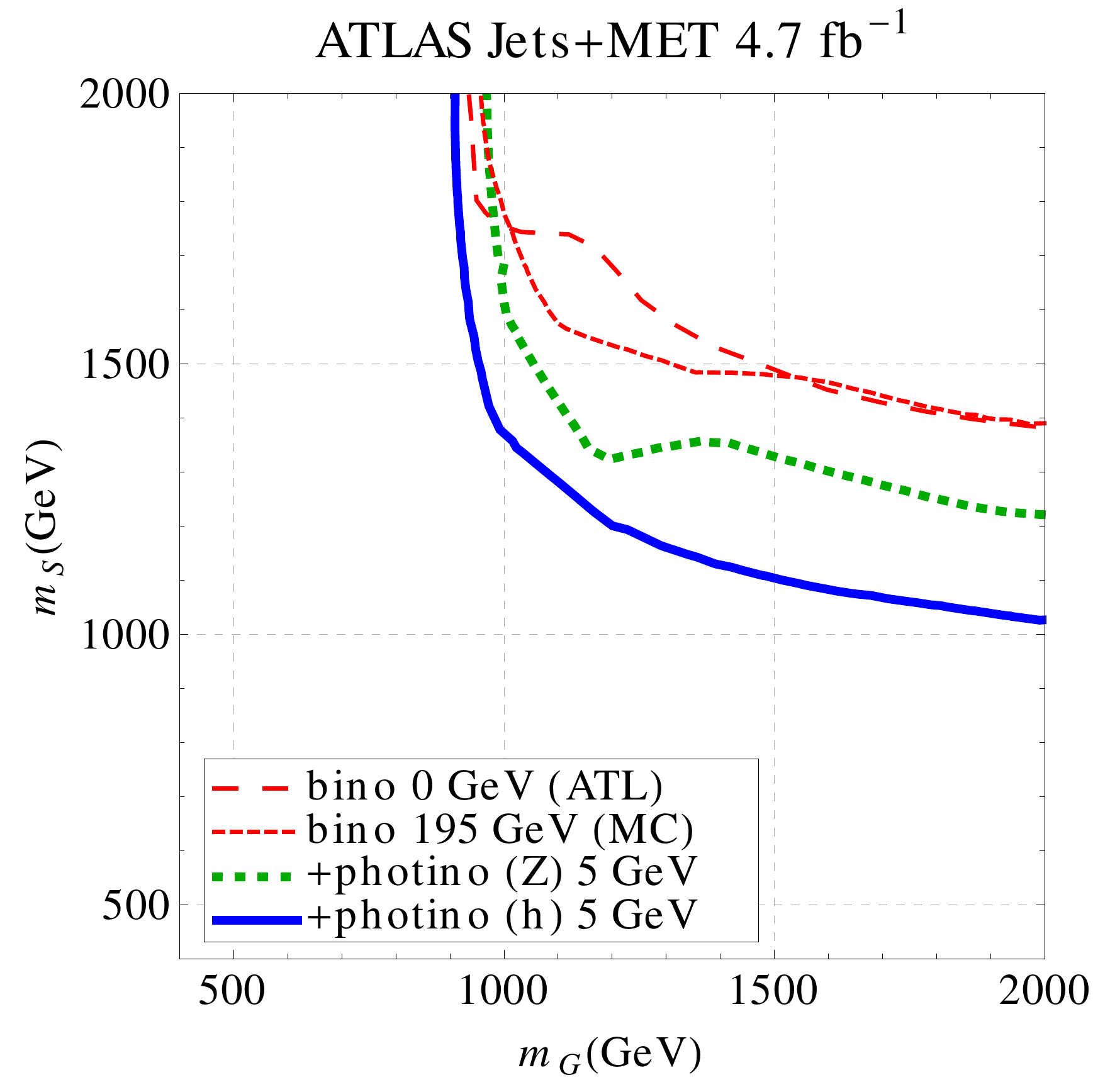}\label{fig:atlasexclusionb}}
    \caption{Allowed parameter space of the gluino-squark-bino
      simplified model for (a)~the ATLAS 2-4 jets+$\slashed{E}_T$
      search at $1.04\ifb$, and (b)~the ATLAS 2-6 jets+$\slashed{E}_T$
      search at $4.7\ifb$.  The red large-dashed and small-dashed lines
      are the 95\% exclusion limits from the ATLAS experiment and from
      our simulation matching, respectively. The green dotted curves
      are exclusion limits based on our scan of spectra with a bino
      LOSP decaying to a single photino through a $Z$ boson, while the
      blue solid curves are the analogous limits for the bino
      decaying via a higgs. } \label{fig:atlasexclusion}
   \end{center}
\end{figure}

We first consider the exclusions from the ATLAS jets+$\slashed{E}_T$
searches in Figure~\ref{fig:atlasexclusion} with the $1.04\ifb$
\cite{ref:atlas2to4jets1fb,ref:atlas2to4jets1fb2} and $4.7\ifb$
\cite{ref:atlas2to6jets5fb} analyses. The ATLAS analysis is presented
for a simplified model of gluino and light-generation squarks decaying
to a LSP. The $1.04\ifb$ analysis includes LSP masses of $0,~195$, and
$395\gev$, while the $4.7\ifb$ analysis only includes direct limits on
a massless LSP. Since in our scenario the bino-like neutralino is the
LOSP and decays to photini, we consider the case of $195\gev$ LSP for
comparison. In the exclusion limits published by ATLAS, there is very
little distinction between the massless and $195\gev$ LSP for the high
mass gluino-squark region, and we find excellent agreement with our
analysis and the experimental exclusion curve with both sets of
data. In the case of the $1.04\ifb$ search
(Fig.~\ref{fig:atlasexclusiona}) we rescaled both signal acceptances
by $70\%$ to match the experimental result; the discrepancy is due
to looser acceptance requirements on jets in our analysis
(Appendix~\ref{app:cuts}). In the $4.7\ifb$ search
(Fig.~\ref{fig:atlasexclusionb}) the only slight discrepancy between
our analysis and experiment is in the signal region with the
highest $m_{\rm eff}$ cut. The relative branching ratios for bino decays
through the higgs or the $Z$ depend on the details of the neutralino
mixing matrix; for simplicity we consider the two limits
$\mathrm{Br}(\tilde{N}_1\rightarrow Z+\tilde{N_5})=1$ and
$\mathrm{Br}(\tilde{N}_1\rightarrow h+\tilde{N_5})=1$ in the context
of the simplified models. We observe a significant reduction in the
limit on the squark mass, and a more modest reduction for the gluino
mass. We also consider the ATLAS multijet analysis
\cite{ref:atlas6to9jets} in this scenario; however, especially as the
gluino limits are not significantly eroded, the limit from the $6$-$9$
jet search is subleading and we do not present it here.

\begin{figure}[t]
\begin{center}
    \subfigure[]{\includegraphics[trim = 0mm 0mm 0mm 0mm, clip, width
      = 0.47\textwidth]{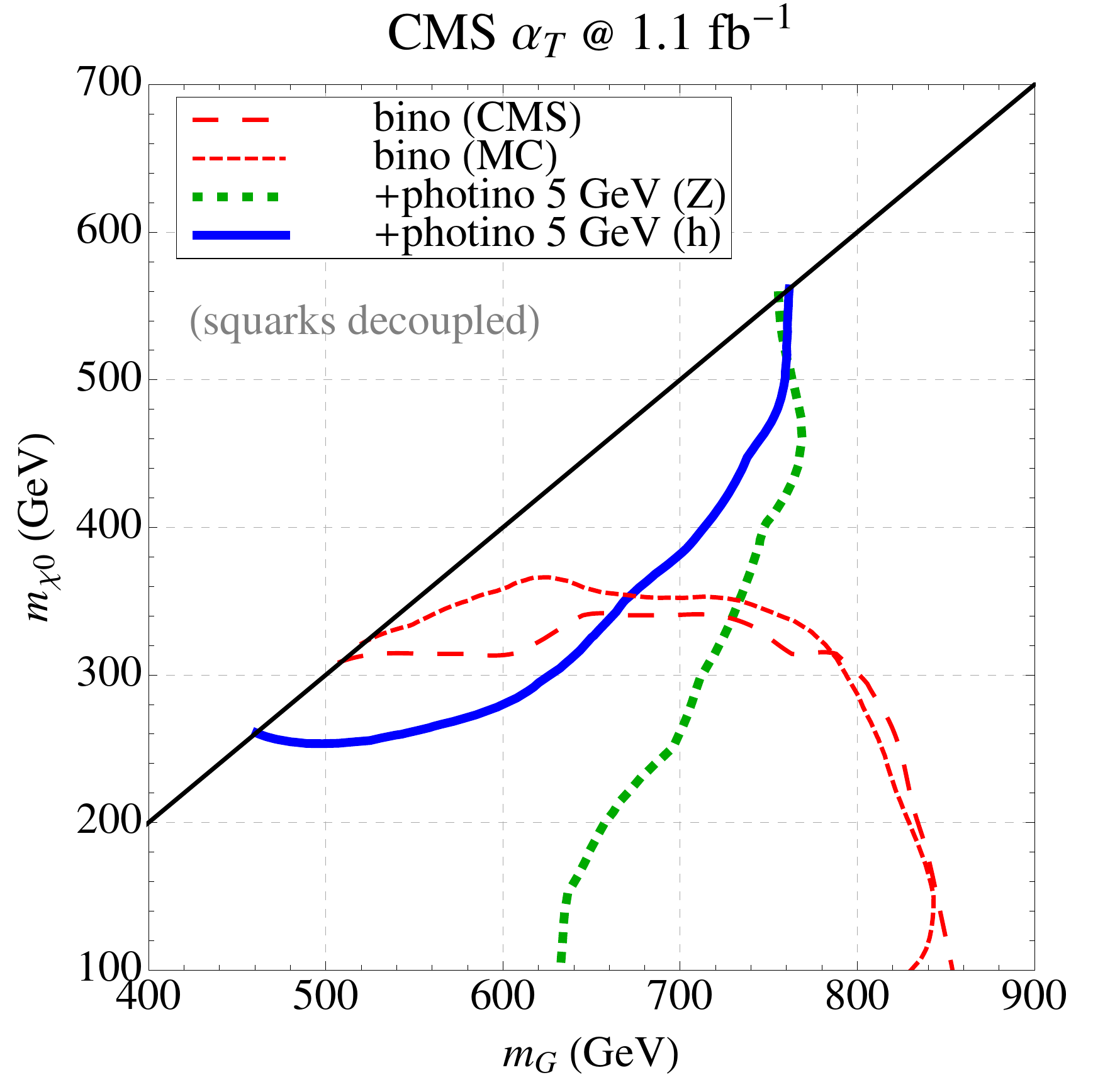}\label{fig:simplifiedexclusiona}}
    \subfigure[]{\includegraphics[trim = 0mm 0mm 0mm 0mm, clip, width
      = 0.47\textwidth]{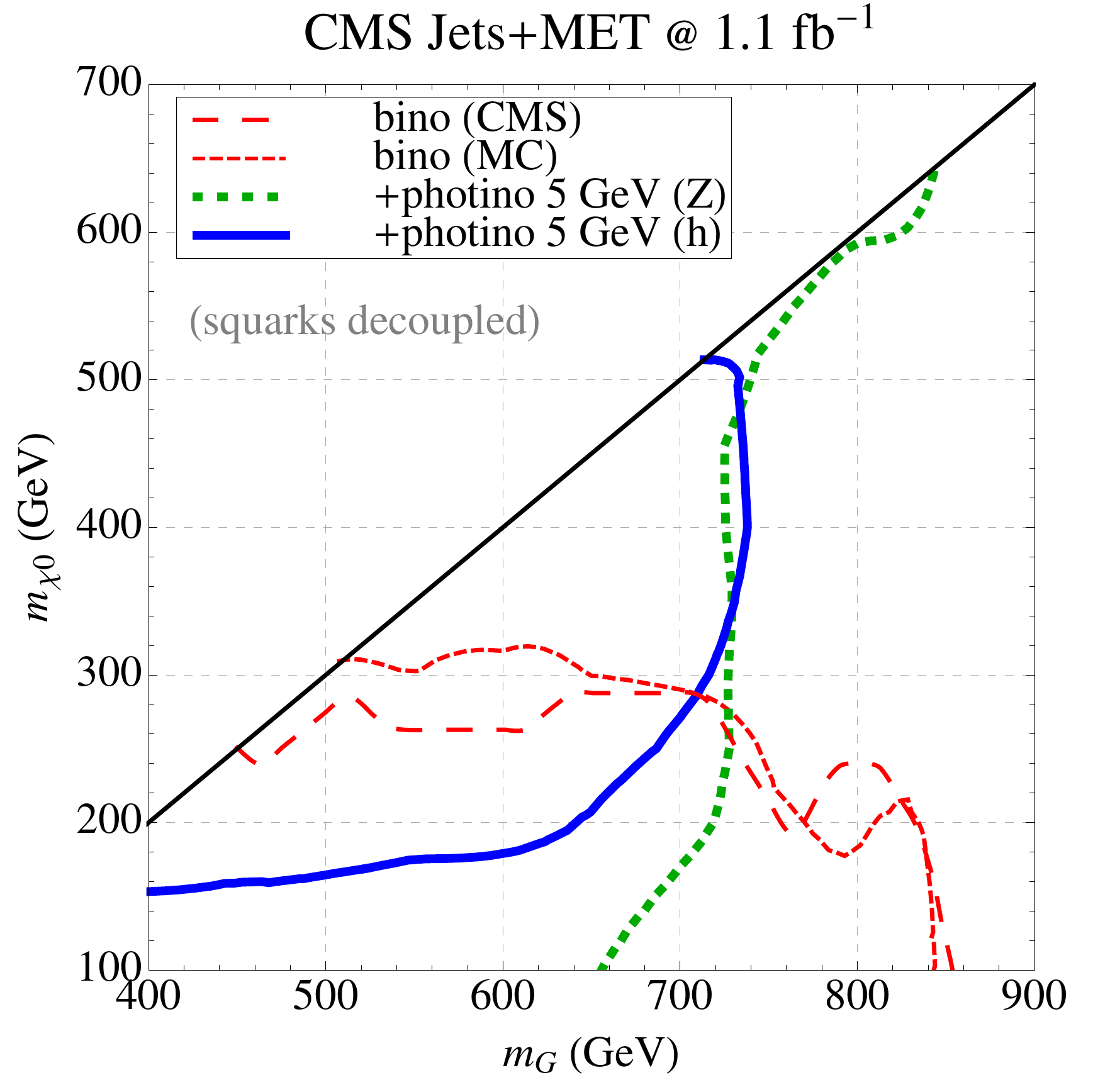}\label{fig:simplifiedexclusionb}}
    \caption{\small{ Allowed parameter space after (a) CMS $\alpha_T$
        and (b) CMS jets+$\slashed{E}_T$ searches at $1
        \ifb$ in the simplified model gluino-bino mass plane. The red
        large-dashed and small-dashed lines are the experimental limit
        and our simulation matching, respectively. The green dotted
        and blue solid lines are the limits corresponding to bino decay via
        $Z$-emission and higgs emission to a photino. Above the black
        line is the kinematic limit of the experimental reach due to
        very high systematic uncertainty for small gluino-bino mass
        splittings; the excluded region is between the diagonal black
        line and the blue (solid) and green (dotted) photino lines. Below LOSP masses
        of $200\gev$, the higgs decay channel exclusion is not
        reliable since large squeezing leads to soft jets and small
        missing energy, causing high sensitivity to the precise
        implementation of the cuts. }}
\label{fig:simplifiedexclusion}
\end{center}
\end{figure}

We present the effect of photini on the CMS $\alpha_T$ and CMS
jets+$\slashed{E}_T$ searches for the gluino-bino simplified models in
Figure~\ref{fig:simplifiedexclusion}.  To best approximate the
experimental procedure, for $\alpha_T$ searches we use a combination
of the $6$ highest $H_T$ signal regions, and for the
jets+$\slashed{E}_T$ search we use the best limit at each point in
parameter space. For heavier LOSPs some of the efficiencies in this
very simple model increase as a result of the extra step in the decay
chain. This is due to two factors: first, in the case of the very
squeezed spectrum of gluino and bino within $300\gev$, the gluino
decay through off-shell squarks produces soft jets, resulting in low
hadronic energy. If the bino then decays hadronically to a much
lighter photino, there is much more visible energy in the
event. Second, the gluinos in the hard collision are produced in
approximately opposite azimuthal directions, so if the mass splitting
with the bino is small, the bino transverse momenta are also nearly
antiparallel and much of the transverse missing energy in the event
cancels. However, if the bino now decays to a $Z$ and photino, the
photino momenta from the two decay chains become uncorrelated and do
not in general cancel, leading to more missing energy. 

On the other hand, with a photino added to the spectrum and a light
bino LOSP, limits on the gluino mass decrease by $200\gev$ or more in
$Z$ decays and reach the kinematic limit in higgs decays. For the
higgs decay, the CMS high $H_T$ signal region sets the tightest
bounds, as missing energy is squeezed out especially for bino masses
near $150 \gev$, close to the higgs at $125\gev$
(Fig~\ref{fig:simplifiedexclusionb}). In the case of the $\alpha_T$
analysis, the search efficiency for low bino-higgs mass splitting is
even lower at $\mathcal{O}(10^{-3})$-$\mathcal{O}(10^{-4})$ as the low
missing energy leads to $\alpha_T$ values of $0.5$ or less, below the
cut for the signal region (Fig~\ref{fig:simplifiedexclusiona}). Below
LOSP masses of $200\gev$, the higgs decay channel exclusion is not reliable
since large squeezing leads to soft jets and small missing energy,
causing high sensitivity to the precise implementation of the cuts.

In Figure~\ref{fig:cmssmexclusion}, we show how the reach for the
hadronic searches (a) CMS $\alpha_T$ and (b) CMS jets+$\slashed{E}_T$
is altered by photini in a cMSSM model framework.  For the $\alpha_T$
search (Figure~\ref{fig:cmssmexclusiona}), the exclusion limit comes
mainly from the highest $H_T$ signal region, and we observe accurate
agreement between the experimental exclusion curve (large dashes) and
the matched exclusion curve on our Monte Carlo data with a bino LSP
(small dashes).  The analogous limit on spectra with photini is
represented by the dashed red curve.  Similarly, in
Figure~\ref{fig:cmssmexclusionb} we show the same curves for the CMS
jets+$\slashed{E}_T$, where teal solid lines represent the limits set
by the ``high $\slashed{H}_T$'' signal region (see Appendix~\ref{app:cuts}
for a more details on the selection cuts).  For both
searches\footnote{We do not present our cMSSM exclusion limit analysis
  for the $\sim5\ifb$ CMS razor and ATLAS 2-6 jets+$\slashed{E}_T$
  SUSY searches.  For the CMS razor search, our matching to the
  experimental limits was poor due to complex statistical methods in
  the CMS analysis, and we found the ATLAS
  jets+$\slashed{E}_T$ limits to be more usefully interpreted in the
  simplified model framework.} ($\alpha_T$ and jets+$\slashed{E}_T$),
we rescaled our Monte Carlo signals for both the bino and photino
samples by a global factor of $75\%$ to match the published
experimental limits for cMSSM spectra.  This rescaling is necessary
due to the treatment of leptons, for which the CMS experiment
generally has looser isolation cuts than \texttt{PGS}.  Since both of
the CMS searches considered here veto on leptons in the event, a
slight overestimate of the signal is to be expected due to the reduced
number of leptons passing isolation cuts.  Note also the slight
differences in shape between our exclusion curves for the bino samples
and the experimental limit curves; this is due to the higher lepton
multiplicity in the region of parameter space with high $m_0$ and low
$m_{1/2}$.\footnote{As a modest validation, this rescaling is not
  necessary to achieve good agreement for the simplified models, where
  the production of leptons is decoupled by construction.}  The
sensitivity of the high $\slashed{H}_T$ signal region is much reduced
for the photino samples.  In fact, for these samples the high $H_T$
channel is the most constraining, although the limits are also reduced
for this signal region.

\begin{figure}[t]
\begin{center}
    \subfigure[]{\includegraphics[trim = 1mm 0mm 0mm 1mm, clip, width
      = 0.48\textwidth]{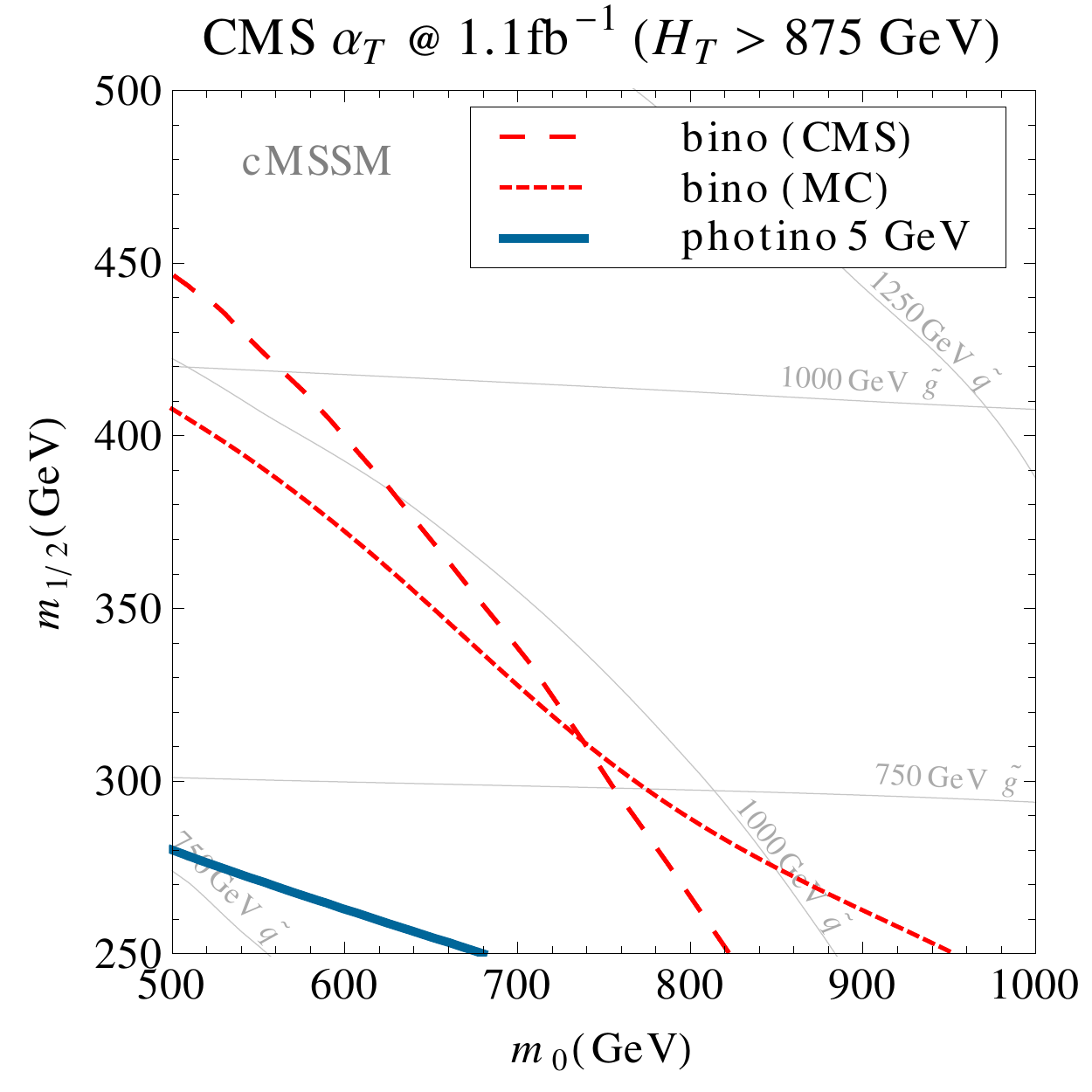}\label{fig:cmssmexclusiona}}
\hspace{-2mm}
    \subfigure[]{\includegraphics[trim = 1mm 0mm 0mm 1mm, clip, width
      = 0.48\textwidth]{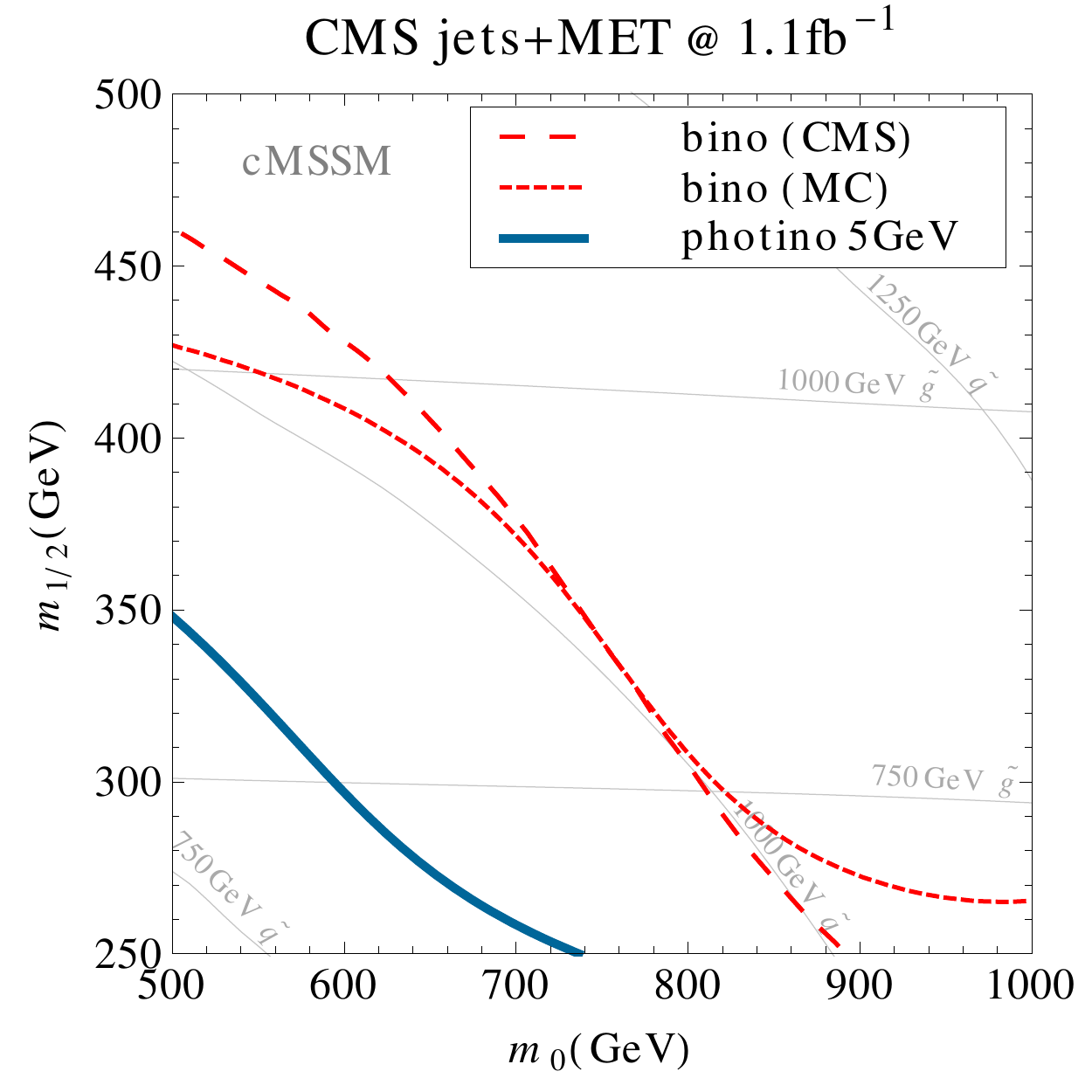}\label{fig:cmssmexclusionb}}
    \caption{Allowed cMSSM parameter space for (a) the CMS $\alpha_T$
      search, and (b) the CMS jets+$\slashed{E}_T$ search.  The red curves
      with large dashing are the 95\% exclusion limits on the cMSSM
      parameter space, as set by the LHC experiments.  The red dashed
      curves are the interpolated exclusion limits based on our scan
      of cMSSM spectra with a bino LSP, while the teal solid curves are the
      analogous limits for the cMSSM spectra with an additional
      photino of 5 GeV. }
\label{fig:cmssmexclusion}
\end{center}
\end{figure}

\section{Finding SUSY}\label{sec:leptonicsearches}

In the previous section, we demonstrated how extending a
supersymmetric spectrum by one or more photini can significantly alter
the nature of experimental signatures and reduce the reach of the
hadronic LHC searches, which currently set the most stringent limits
on SUSY production. However, the decays of the LOSP (taken here to be a mostly-bino
neutralino) to photini via $Z$ and/or higgs bosons can lead to
signatures in other channels, most notably those involving same-sign
(SS) leptons, opposite-sign (OS) leptons, and bottom
quarks, in addition to some missing
energy. The additional contributions to these signal channels may improve their effectiveness in setting limits, and indeed may render them the most promising channels for discovery. We investigate the reach of leptonic SUSY searches at the LHC in two separate frameworks, namely the cMSSM and the gluino-squark-bino simplified model of Section~\ref{sec:simdetails}.

\begin{figure} [t]
  \begin{center}
      \subfigure[]{\includegraphics[trim = 0mm 0mm 0mm 0mm, clip, width
      = 0.48\textwidth]{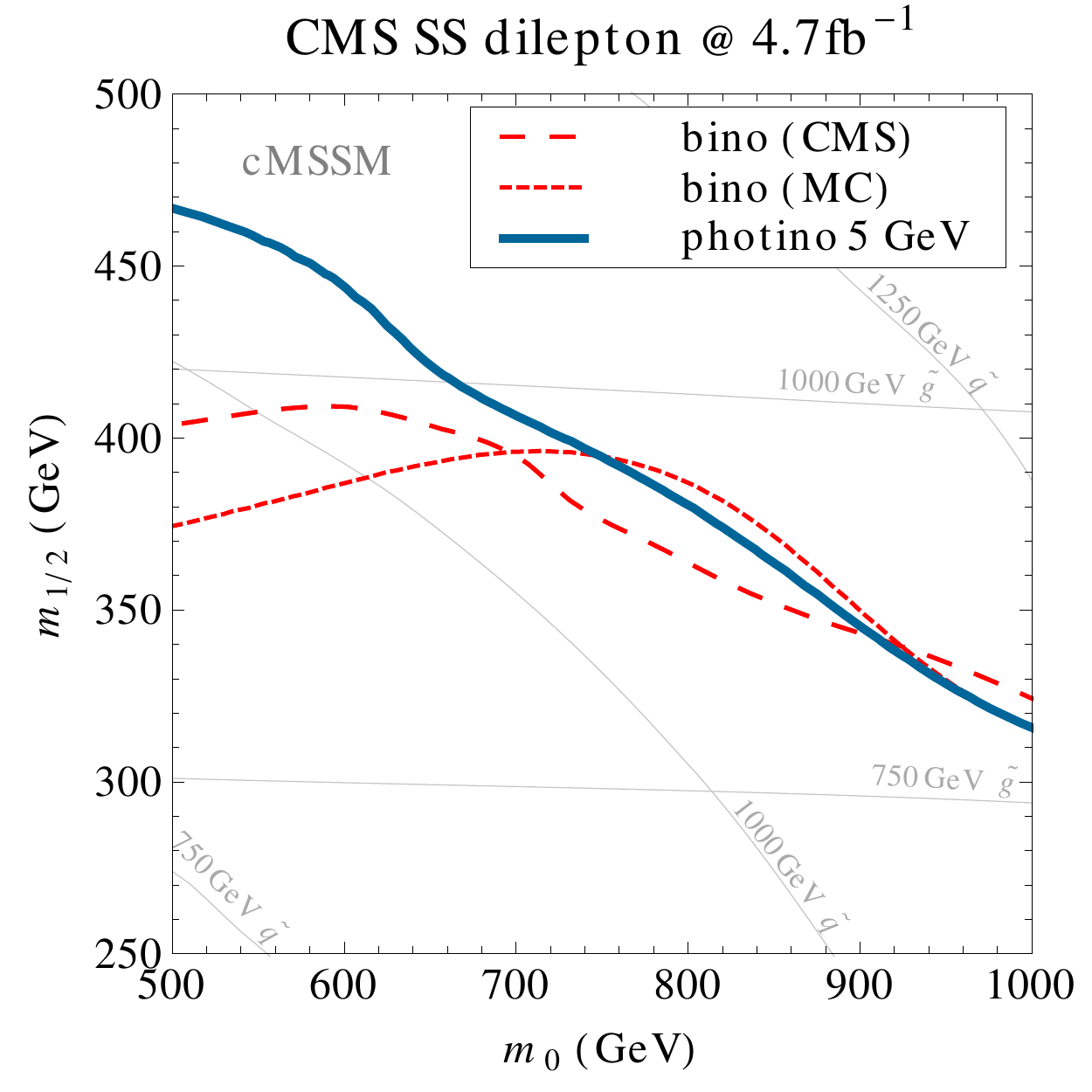}\label{fig:SSeff}}
    \subfigure[]{\includegraphics[trim = 0mm 0mm 0mm 0mm, clip, width
      = 0.48\textwidth]{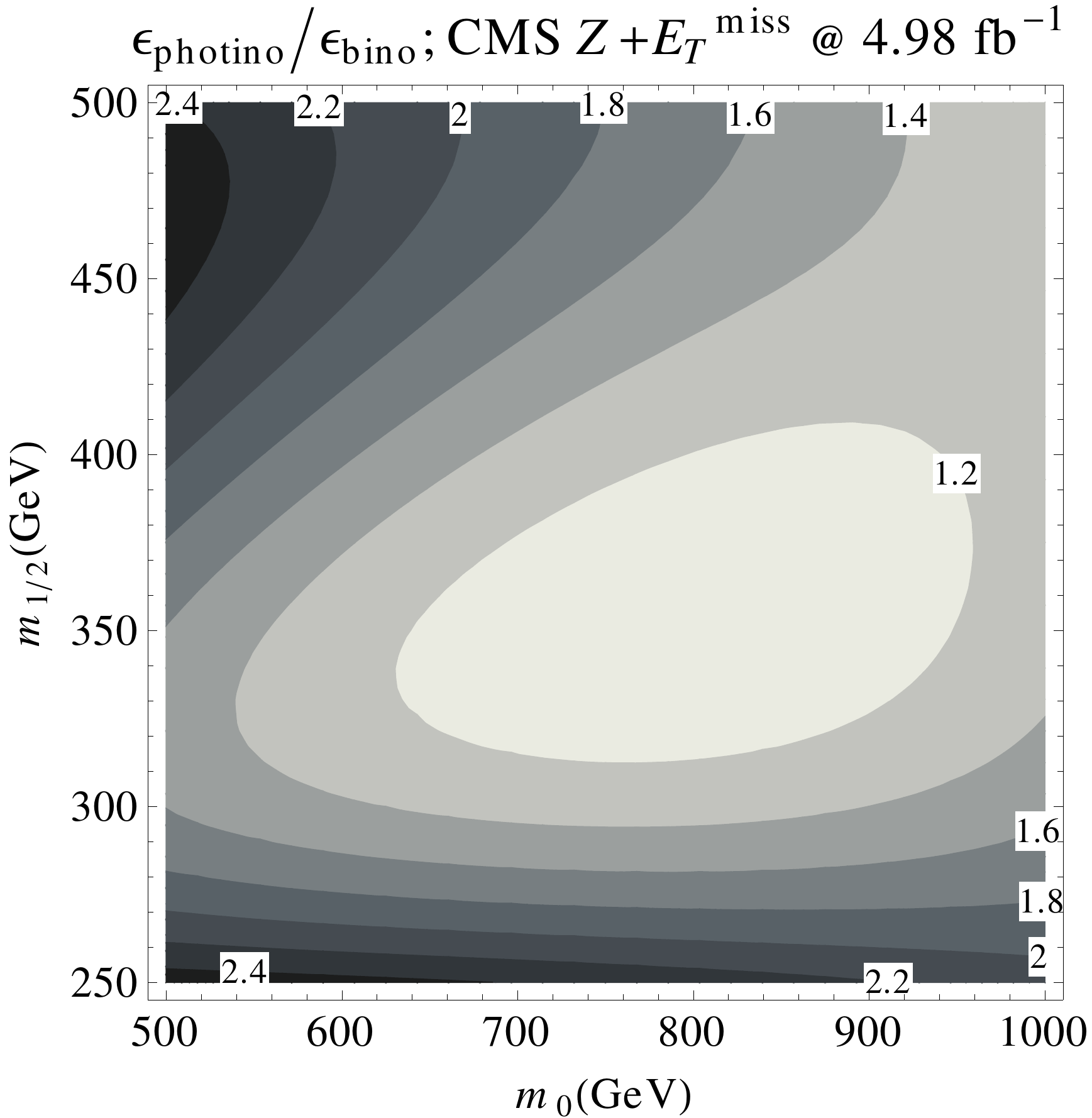}\label{fig:Zeff}} 
 \end{center}
\vspace{-4mm}
\caption[]{\small{(a) The 95\% CL exclusion limit on the cMSSM
    parameter space set by the CMS SS dilepton search using the most
    constraining signal region.  The red large-dashed curve is the limit
    set by CMS, the red dashed curve is the 95\% CL exclusion limit on our
    Monte Carlo data with the bino as the LSP, and the teal solid curve is
    the analogous curve with photino as the LSP. (b) Contours of the
    ratio of the selection efficiency with ($\epsilon_{\text{photino}}$) and without ($\epsilon_{\text{bino}}$) a 5 GeV photino
    in the cMSSM spectrum, for the $Z$+jets+$\slashed{E}_T$ search
    using the $JZB$ variable.}}
\label{fig:LeptonEff}
\end{figure}

To study the sensitivity of leptonic searches in the cMSSM framework,
we focus on the dedicated CMS searches for SS dileptons and
$\slashed{E}_T$ \cite{ref:SS5fb,ref:SS1fb}, and leptonically-decaying
$Z$ bosons and $\slashed{E}_T$ \cite{ref:ZJetsmet}. We perform simple
versions of the CMS SS dilepton search at 4.7 $\ifb$ and the
$Z$+jets+$\slashed{E}_T$ SUSY search at $4.98\ifb$ on our cMSSM Monte
Carlo samples to estimate the leptonic detection reach for our model.
The SS dilepton search looks for sufficiently hard pairs of $ee$,
$e\mu$ or $\mu\mu$ along with a minimum $\slashed{E}_T$ requirement
and a veto on opposite-sign, same-flavor lepton pairs that reconstruct
the $Z$.  The $Z$+jets+$\slashed{E}_T$ search selects events based on
jet energy, lepton energy, and jet-$Z$ balance (dubbed $JZB$) with
cuts similar to those listed in \cite{ref:ZJetsmet}.  The $JZB$
variable can be thought of as a measure of $\slashed{E}_T$ with sign
information; under- or over-measurement of the jet energy generally
leads to negative or positive $JZB$, respectively.  Since jet energy
is more likely to be under-measured in a detector and SUSY events
typically have signatures with positive $JZB$, cutting on events with
the requirement $JZB>150\gev$ is an effective way to achieve good
signal versus background separation.  More detailed criteria for both
searches are listed in Appendix~\ref{app:cuts}.

In Figure~\ref{fig:SSeff}, we show the limits on the cMSSM set by the
CMS SS dilepton search in the most constraining signal region (SR \#4,
see Appendix~\ref{app:cuts}). To account for the discrepancy
of lepton detection efficiencies between our PGS simulation and
experiment, we rescale our efficiencies by $80\%$.  We can see that the detection reach is
largely unaffected by photini despite the $\slashed{E}_T$ reduction
and a higher occurrence of $Z$ vetos from OSSF leptons, as there is
also a higher multiplicity of SS lepton pairs. As such, the SS
dilepton search remains an effective means of looking for SUSY with
photini cascades, and in many cases may set the most stringent limits
on the spectrum. This raises the tantalizing possibility of first
discovering SUSY in a leptonic search. 

In Figure~\ref{fig:Zeff}, we show a contour plot of the ratio of
selection efficiencies for the $Z$+jets+$\slashed{E}_T$ search with
and without a photino of 5 GeV. We present selection efficiencies only
because CMS has not yet released limits on the full cMSSM parameter
space in this search. However, any systematic errors in our treatment
of leptons will largely cancel out in a ratio of efficiencies, and
suffices to paint a qualitative picture of the change in detection
reach in the presence of photini.  We observe that the sensitivity of
the search increases by a factor of 1.2 to 2.4, depending on the
spectrum and the signal region. The contours shown are for the signal
region with the tightest $JZB$ requirements; for other signal regions
with lower $JZB$ cuts, the ratio of bino to photino efficiencies
changes by up to $10\%$. Unsurprisingly, this suggests an increased
sensitivity of the $Z$+jets+$\slashed{E}_T$ search to cascade decays
involving light photini due to the additional on-shell $Z$ bosons.

There are other searches that may prove to be sensitive to
detecting extended SUSY cascades with photini, though for various reasons they
are typically somewhat less effective than SS dileptons when applied to the cMSSM. 
The multilepton searches \cite{ref:multileptons} are not particularly sensitive to OSSF pairs
of leptons with invariant mass near the $Z$ mass, but may prove
constraining when there is additional leptonic activity in the decay
chain or when decays proceed predominantly through the higgs.  Note
that the multilepton search becomes more effective when the mass
splitting between the LOSP and the lightest photino is small and the
$Z$ is far off-shell, but in this case the $\slashed{E}_T$
distribution is not significantly altered and standard hadronic SUSY
searches provide stronger limits. Ancillary searches such as
opposite-sign dilepton plus $\slashed{E}_T$ \cite{ref:ZJetsoffshell}
mainly look for a kinematic edge of OSSF leptons outside of the
di-lepton invariant mass window of the $Z$ boson, rendering these
studies less effective for constraining the case at hand.  Searches
involving bottom quarks \cite{ref:bjets1,ref:bjets2,ref:bjets3} may
become sensitive especially when the dominant decay from the bino to a
photini occurs through emission of a light higgs boson, though for
these decays the $\slashed{E}_T$ signal is significantly
reduced.

\begin{figure} [t]
  \begin{center}
      \subfigure[]{\includegraphics[trim = 0mm 0mm 0mm 0mm, clip, width
      = 0.48\textwidth]{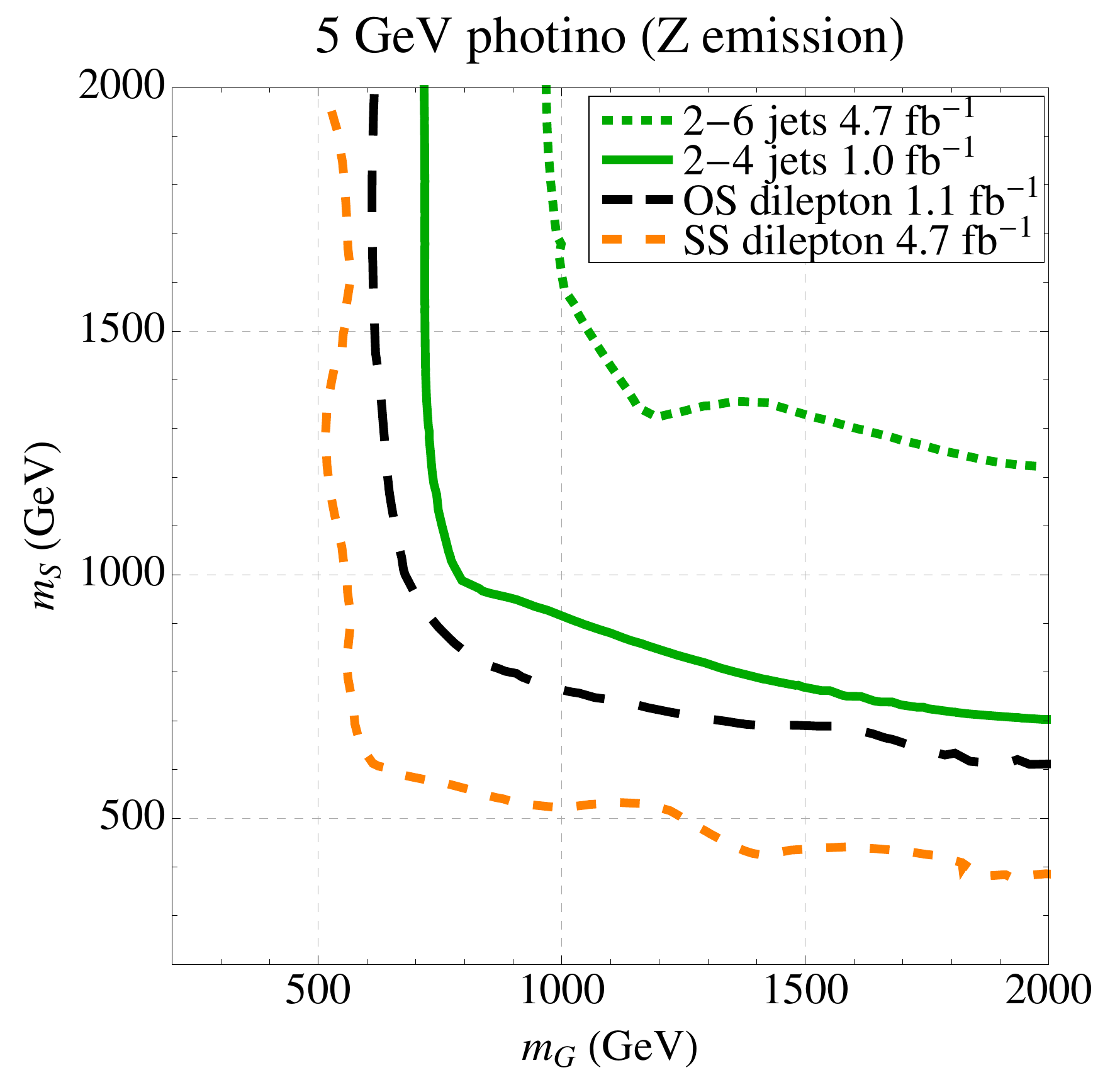}}
    \subfigure[]{\includegraphics[trim = 0mm 0mm 0mm 0mm, clip, width
      = 0.48\textwidth]{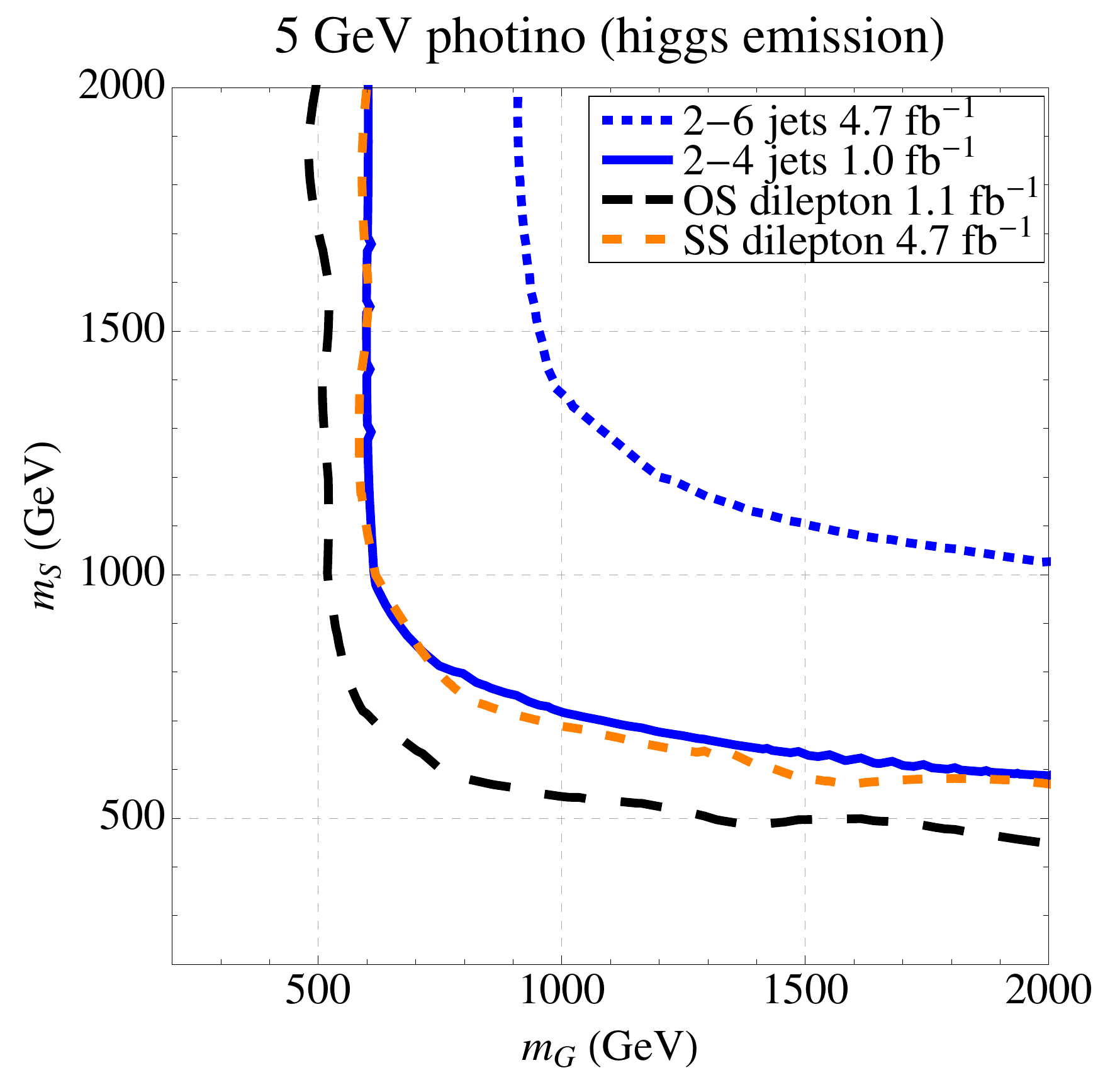}} 
 \end{center}
\vspace{-4mm}
\caption[]{\small{ Exclusion limits on the squark-gluino-bino simplified
    model in the presence of a $5\gev$ photino in the case the LOSP
    decays to the photino via (a) a $Z$ boson or via (b) a higgs
    boson. We take only the first two squark generations light and the
    remaining sparticles decoupled at $3.5\tev$. We compare the reach
    of the ATLAS 2-6 jets and $\slashed{E}_T$ search for
    (a) $Z$ emission (dotted, green curve) and (b) higgs emission (dotted, blue curve), the ATLAS 2-4 jets and
    $\slashed{E}_T$ search for (a) $Z$ emission (solid, green curve) and (b) higgs emission (solid, blue curve), to two leptonic searches: OS dileptons (black,
    large dashed), and SS dileptons (orange, small dashed). The OS
    dilepton search has a larger missing transverse momentum cut,
    leading to a weaker limit on higgs emission in (b) compared to $Z$ emission in (a).} }
\label{fig:SimpleLepReach}
\end{figure}

For completeness, we also consider the implications of leptonic searches for our simplified models, though in this case the decoupling of sleptons and electroweakinos has an important impact on the relative sensitivity of various searches. 
Focusing on the squark-gluino-bino simplified model, we perform analyses
similar to the CMS same-sign dilepton search (with $5\ifb$,
\cite{ref:SS5fb}) and ATLAS opposite-sign dilepton search (with
$1\ifb$, \cite{ref:atlasOS1fb}), and compare to the reach of the
hadronic ATLAS jets searches for $1\ifb$ and $5\ifb$ considered in
Section~\ref{sec:hadronicsearches}.  A simplified spectrum in which
only squarks, gluinos, and binos are accessible at LHC energies does
not yield many leptons by construction, so here we only consider the
case of the bino decaying to a photino by $Z$ or higgs boson emission
(and not the case of a stable bino).  There is an additional
uncertainty in our results for these two leptonic searches from our
treatment of leptons; in this simplified model framework we could not
validate our exclusion curves against published experimental exclusion
curves with the bino as the LSP (as opposed to the cMSSM framework,
e.g.~see the two red curves in Figure~\ref{fig:SSeff}). We implement a
rescaling of the efficiencies by $80\%$ as in the cMSSM scenario to
account for experimental leptonic efficiencies.
 
As the photini cascades inject additional leptons into the events
through $Z$ or higgs boson emission, the leptonic searches become
competitive with jets and missing energy searches for these simplified
models. In Figure~\ref{fig:SimpleLepReach} we demonstrate that even in
the simplified model context, the opposite-sign lepton with missing
energy searches are comparable to the 2-4 jets and $\slashed{E}_T$
search with the same integrated luminosity. In the case of a more
complete supersymmetric spectrum, we expect the leptonic searches to
be even more effective, as the decays to photini will be one of
several sources of leptons (in addition to slepton and stop decays,
for instance). This is particularly true in the case of SS dilepton
searches, which are very effective for the cMSSM (see
Figure~\ref{fig:SSeff}) but less so for simplified models; to pass the
SS dilepton cuts as shown in Figure~\ref{fig:SimpleLepReach}, both
$Z$'s or higgses in the event have to decay leptonically, reducing the
efficiency by small branching ratios. Other leptonic searches
such as CMS multileptons \cite{ref:multileptons} or ATLAS trileptons
\cite{ref:atlastrileptons} could be very sensitive to photini produced
via higgs decays in the $Z$-veto signal regions.  However, we note
that the current ATLAS trilepton search also vetoes on b-tags, which
nearly eliminates the signal. This underlines the importance of
leptonic searches with 0, 1, or 2 b-tags in order to capture higgs
decays in supersymmetric decay chains.

The sensitivity of the searches considered here highlights the fact that the presence of additional gauge fermions mixing with the bino may mitigate limits from the hadronic SUSY searches with the greatest reach, but does not extinguish SUSY signals entirely. Indeed, it suggests that in many cases the first SUSY signals may arise in search channels involving $Z$'s, leptons, or other supplementary tags whose sensitivity is unaltered or improved by the presence of photini. Finally, it would be especially interesting to probe the
sensitivity of proposed SUSY searches that rely on lepton and jet
multiplicity, rather than $\slashed{E}_T$ \cite{Lisanti:2011tm}, since
extended cascade decays involving photini are likely to enhance the
relevant SUSY signal.

\section{Conclusions}

In this work we have studied the phenomenology of photini --
additional light abelian gauge fermions -- on SUSY searches at the
LHC.
  The existence of these light fermionic degrees of freedom is
natural in the context of string compactifications, though the
phenomenology is common to any models in which extra neutral fermions
mix with MSSM neutralinos. We performed a detailed analysis of the
impact of SUSY spectra with photini on current hadronic searches; the
limits set on squark and gluino masses by the most constraining
searches at both ATLAS and CMS are reduced dramatically.  The
reduction persists for spectra in two simplified models as well as the
cMSSM, with the latter case providing a reasonable sample of longer
cascade decays.

In particular, for a simplified model consisting of a gluino,
light-flavor squarks, and a $195\gev$ bino LOSP, bounds on the squark
masses are relaxed to as low as $1\tev$ from the current best limit of
$1.4\tev$ squarks for a $2\tev$ gluino. While ``squeezed'' spectra
with a bino-gluino splitting below $300\gev$ actually become more
constrained in the presence of photini due to kinematics, the limit
reduction is dramatic for natural LOSP masses of the same order as the
higgs mass.  In fact, in the limit in which all sparticles are
decoupled except for the gluino and LOSP (motivated by e.g. split
supersymmetry), the leading hadronic SUSY searches at CMS cannot
exclude any spectrum with gluinos heavier than $400\gev$ and a bino of
mass $m_{\tilde{N_1}}\lesssim 160\gev$ that dominantly decays to the
lightest photino through $125\gev$ higgs emission. The CMS $\alpha_T$
variable in particular is only weakly sensitive to photini events with
small missing energy.  More standard hadronic and $\slashed{E}_T$
searches at ATLAS and CMS are less compromised, but set reduced limits
up to the kinematic edge of a LOSP nearly degenerate in mass with the
higgs.

We emphasize that we have only considered one limit of photino
parameter space, in which photini couplings are moderate and decays
proceed through on-shell Standard Model fields. There are a variety of
other parametric regions that may reduce the sensitivity of SUSY
searches. In particular, as the photini mixing is reduced, decays
involving photini may involve displaced vertices. Depending on the
decay length, these cascades may evade searches that depend on impact
parameter and track quality -- a sort of $R$-preserving version of
displaced supersymmetry \cite{Graham:2012th}.

In general, supersymmetry concealed by photini or similar light degrees of freedom
will not be impossible to discover at the LHC. Although the
sensitivity of frontier searches relying on jets and missing energy
may be significantly eroded, it will not be entirely
compromised. Ancillary searches with sensitivity to $Z$'s or
additional leptons should provide compensatory coverage, although
their current reach is not as great. For the most part, exceptionally
sophisticated new techniques will not be required, but rather the
persistent pursuit of a wide range of complementary SUSY search
strategies. But by eroding the sensitivity of the SUSY searches with
greatest reach, the presence of photini in decay chains may render
natural supersymmetric spectra more compatible with current LHC data
and increase the prospects of discovering supersymmetry in novel
channels.

\section*{Acknowledgments}
\noindent We thank Asimina Arvanitaki, Erik Devetak, Savas Dimopoulos,
Eleanor Dobson, Sergei Dubovsky, and John March-Russell for useful
conversations, and Josh Ruderman and Prashant Saraswat for helpful comments on
the manuscript. We are especially grateful to Eder Izaguirre and Jay
Wacker for providing Madgraph simplified model data, and to Benjamin
Fuks for support regarding the Feynrules program.  This work was
supported in part by ERC grant BSMOXFORD no. 228169.  MB is supported
in part by the NSF Graduate Research Fellowship under Grant
No. DGE-1147470.  NC is supported by NSF grant PHY-0907744, DOE grant
DE-FG02-96ER40959, and the Institute for Advanced Study, and
acknowledges hospitality from the Stanford Institute for Theoretical
Physics where parts of this work were completed.

%\clearpage
\appendix
\section{Selection cuts}\label{app:cuts}
\begin{enumerate}
 \item CMS $\alpha_T$ search ($1.1 ~\text{fb}^{-1}$) \cite{ref:alphaT1fb}:
    \begin{itemize}
      \item jet requirements of $E_T > 50~\gev$ and pseudorapidity $|\eta|<3$ 
      \item $H_T > 275~\gev$, where $H_T$ is defined as the scalar sum
        of the jet $E_T$'s; we combine limits from the following 6
        $H_T$ bins throughout our paper:
        ($375$-$475,475$-$575,575$-$675,675$-$775,775$-$875,
        875+)$;\footnote{In addition, \cite{ref:alphaT1fb} defines two
          lower $H_T$ bins, $(275$-$325)$ and $(325$-$375)$ but these
          set no limits for the spectra considered in our paper.}
      \item lepton veto (no electron or muon with $p_T>10~\gev$);
      \item central leading jet (the hardest jet needs to have $|\eta|<2.5$;
      \item two hard jets (the hardest two jets both need to have $E_T > 100~\gev$);
      \item $R_{\text{miss}}\equiv \frac{\slashed{H}_T}{\slashed{E}_T}<1.25$, where the $\slashed{H}_T$ is a jet-based measure of missing energy (namely the $E_T$ of the vectorial sum of all jets) and $\slashed{E}_T$ is the missing transverse energy measured at the calorimeter level
      \item $\alpha_T \equiv \frac{E_T^{\text{j2}}}{M_T}> 0.55$, where $M_T$ is defined as the total transverse mass of all jets, and $E_T^{\text{j2}}$ is the transverse energy of the sub-leading jet if the jet multiplicity $n=2$ (for $n>2$, two pseudojets are first formed according to \cite{ref:alphaT1fb}).
    \end{itemize}

 \item CMS razor variables search ($ 4.4~\text{fb}^{-1}$)
   \cite{ref:razor5fb}:\footnote{We implement these cuts for
     Fig.~\ref{fig:razor} to illustrate the shifts in the $M_R$ and $R^2$ variables. The full search contains a more baseline requirements, such as a veto on leptons.}
    \begin{itemize}
      \item jets need to have $E_T>60\gev$ and $|\eta|<3$;
      \item $1000~\gev< M_R < 3500~\gev$ and $0.18 < R^2 < 0.5$, where $M_R \equiv \sqrt{(E^{\text{j}_1}+E^{\text{j}_2})^2 - (p_z^{\text{j}_1}+p_z^{\text{j}_2})^2 }$, 
      $M^R_T \equiv \sqrt{\frac{\slashed{E}_T(p_T^{\text{j}_1}+p_T^{\text{j}_2}) - \vector{\slashed{E}}_T(\vector{p}_T^{\text{j}_1}+\vector{p}_T^{\text{j}_2})}{2}}$, $R\equiv \frac{M^R_T}{M_R}$, and $\text{j}_1$/$\text{j}_2$ are two ``megajets'' defined according to \cite{ref:razor5fb}.
    \end{itemize}
    
 \item CMS jets+$\slashed{E}_T$ search ($ 1.1~\text{fb}^{-1}$)\cite{ref:cmsjetsmet1fb}:
    \begin{itemize}
      \item $\ge 3$ jets with $p_T>50\gev$ and $|\eta|<2.5$;
      \item lepton veto (no electron or muon with $p_T>8\gev$ and $|\eta|<2.1$));
      \item $H_T > 350\gev$, where $H_T$ is calculated based on jets with $p_T > 50\gev$ and $|\eta|<5$;
      \item $\slashed{H}_T > 200\gev$, where $\slashed{H}_T$ is
        calculated based on jets with $p_T > 30\gev$ and $|\eta|<5$;
      \item $\slashed{H}_T$ four-momentum isolation from the three leading jets in the $eta-\phi$ plane (by a distance of 0.5 for the two leading jets, 0.3 for the third-hardest jet)
      \item three signal regions: ``high $H_T$'' ($H_T > 800\gev)$, ``medium $H_T$ and $\slashed{H}_T$'' ($H_T>500\gev$, $\slashed{H}_T>350\gev$), and ``high
          $\slashed{H}_T$'' ($\slashed{H}_T>500\gev$) as described in \cite{ref:cmsjetsmet1fb}.
    \end{itemize}
    
 \item ATLAS $2$-$4$ jets+$\slashed{E}_T$ search ($1.04~\text{fb}^{-1}$) \cite{ref:atlas2to4jets1fb}:
 \begin{itemize}    
   \item $\slashed{E}_T > 130$ GeV 
       \item Leading jet $p_T>130$ GeV, other jets $p_T>80$ GeV, for
         5 or more $p_T>40$ GeV,
       \item $m_{\text{eff}}$ cuts between $500$ and $1100$ GeV where
         $m_{\text{eff}} = \slashed{E}_T +\sum_{\text{jets}} |p_T|$,
       \item lepton veto (no electron or muon with $p_T>20\gev$),
       \item $\slashed{E}_T /m_{\text{eff}} $ cuts between $0.2$ and $0.3$,
       \item Five signal regions based on number of jets,
         $m_{\text{eff}}$, and $\slashed{E}_T /m_{\text{eff}} $ as in \cite{ref:atlas2to4jets1fb}. The
         limit on signal is set by the most constraining region at
         each point in parameter space.
 \end{itemize}
 
 \item ATLAS $2$-$6$ jets+$\slashed{E}_T$ search ($4.7~   \text{fb}^{-1}$) \cite{ref:atlas2to6jets5fb}
 \begin{itemize}    
       \item $\slashed{E}_T > 160$ GeV;
       \item leading jet $p_T>130$ GeV, other jets $p_T>60$ GeV, for
         5 or more $p_T>40$ GeV;
       \item $m_{\text{eff}}$ cuts between $900$ and $1900$ GeV where
         $m_{\text{eff}} = \slashed{E}_T +\sum_{\text{jets}} |p_T|$;
       \item lepton veto (no electron or muon with $p_T>20~\gev$);
       \item $\slashed{E}_T /m_{\text{eff}} $ cuts between 0.15 and 0.4;
       \item  11 signal regions based on number of jets,
         $m_{\text{eff}}$, and $\slashed{E}_T /m_{\text{eff}} $ as in \cite{ref:atlas2to6jets5fb}. The
         limit on signal is set by the most constraining region at
         each point in parameter space.
 \end{itemize}
 
 \item CMS same sign (SS) dilepton search ($4.7 \ifb$) \cite{ref:SS5fb,ref:SS1fb}, high-$p_T$
   lepton baseline, signal region \#4
 \begin{itemize}
  \item $\ge 2$ jets in event, each with $p_T > 40\gev$ and $|\eta|<2.5$;
  \item baseline lepton requirements of $p_T>5\gev$ for muons, $p_T>10\gev$ for electrons, $|\eta|<2.4$ for both flavors;
  \item  veto on opposite sign same flavor (OSSF) pairs that reconstruct the $Z$ (i.e.~$76\gev < m_{ll} < 106\gev$);
  \item at least one same-sign (SS) dilepton pair of type $e\mu$, $ee$ or $\mu\mu$, each with $p_T>10\gev$;
  \item the leading lepton in the event (not necessarily out of the dilepton pair) has $p_T>20\gev$;
  \item $\slashed{E}_T>100\gev$.
 \end{itemize}

 \item $JZB$+jets+$\slashed{E}_T$ \cite{ref:ZJetsmet}
       \begin{itemize}
       \item $\ge 3$ central jets with $p_T > 50~\gev$ and $|\eta|<2.5$;
       \item at least one pair of OSSF leptons with $p_T > 20\gev$ which reconstruct the $Z$ boson to mass to within 20 GeV ($|m_{l^+l^-}-m_Z|<20\gev$);
       \item the hardest isolated OSSF lepton pair needs to have $JZB > 150\gev$, where $JZB$ is defined as
  \begin{equation} JZB \equiv \left| \sum_{\text{jets}} \vector{p}_T \right| -  \left| \vector{p}^{(Z)}_T \right| \end{equation}
  and serves as a measure of the $\slashed{E}_T$ in the event.
\end{itemize}

 \item ATLAS dilepton and missing transverse momentum search ($1 \ifb$)
   \cite{ref:atlasOS1fb}, ``OS-inc'' signal region
 \begin{itemize}
  \item $E_T^{\mathrm{miss}} > 250\gev$, where $E_T^{\mathrm{miss}} $
    is the magnitude of the vector sum of the $p_T$ of jets with $p_T
    >20\gev$ and signal leptons;
  \item baseline lepton requirements of $p_T>20\gev,~|\eta|<2.47$ for
    electrons, $p_T>10\gev,~|\eta|<2.4$ for muons;
 \item  isolation requirements on leptons: energy within a cone of
   $\Delta R < 0.2$ is less than $10\%$ of the $p_T$ for electrons or
   $1.8\gev$ for muons;
  \item  exactly two leptons as defined by the above momentum and
    isolation cuts in the event;
  \item  invariant dilepton mass $m_{ll}>12\gev$ to suppress low-mass resonances;
  \item one opposite-sign (OS) dilepton pair of type $e\mu$, $ee$ or $\mu\mu$;
  \item the leading lepton in the event has $p_T>20\gev$ for a muon or
    $p_T>25\gev$ for an electron.
\end{itemize}

\end{enumerate}

\bibliography{photinibib}
\bibliographystyle{JHEP}

\end{document}